\documentclass[conference]{IEEEtran}
\IEEEoverridecommandlockouts
\usepackage{cite}
\usepackage{amsmath,amssymb,amsfonts}
\usepackage{algorithmic}
\usepackage{graphicx}
\usepackage{textcomp}
\usepackage{xcolor}

\usepackage{algorithm}
\usepackage{cases}
\usepackage{subfigure}
\usepackage{graphicx}
\usepackage{epsfig}
\usepackage{mathrsfs}
\newtheorem{defn}{Definition}[section]
\usepackage{color}
\usepackage{fancyhdr}
\usepackage{multirow}
\newtheorem{theorem}{Theorem}
\newtheorem{lemma}[theorem]{Lemma}

\def\BibTeX{{\rm B\kern-.05em{\sc i\kern-.025em b}\kern-.08em
    T\kern-.1667em\lower.7ex\hbox{E}\kern-.125emX}}
\begin{document}

\title{Data Provenance via Differential Auditing}

\author{\IEEEauthorblockN{Xin Mu}
\IEEEauthorblockA{Peng Cheng Laboratory \\
 Shenzhen, China \\
 mux@pcl.ac.cn}
 \and
 \IEEEauthorblockN{Ming Pang}
 \IEEEauthorblockA{JD, China \\
 pangm@jd.com}
\and
 \IEEEauthorblockN{Feida Zhu
}
 \IEEEauthorblockA{Singapore Management University\\
 Singapore\\
fdzhu@smu.edu.sg}
 }

\maketitle

\begin{abstract}
Auditing Data Provenance (\emph{ADP}), i.e., auditing if a certain piece of data has been used to train a machine learning model, is an important problem in data provenance. The feasibility of the task has been demonstrated by existing auditing techniques, e.g., shadow auditing methods, under certain conditions such as the availability of label information and the knowledge of training protocols for the target model. Unfortunately, both of these conditions are often unavailable in real applications. In this paper, we introduce Data Provenance via Differential Auditing (\textsf{DPDA}), a practical framework for auditing data provenance with a different approach based on statistically significant differentials, i.e., after carefully designed transformation, perturbed input data from the target model's training set would result in much more drastic changes in the output than those from the model's non-training set. This framework allows auditors to distinguish training data from non-training ones without the need of training any shadow models with the help of labeled output data. Furthermore, we propose two effective auditing function implementations, an additive one and a multiplicative one. We report evaluations on real-world data sets demonstrating the effectiveness of our proposed auditing technique. 
\end{abstract}


\section{Introduction}
In an era of accelerated digital transformation, data has been widely recognized as an emerging asset class. Data provenance \cite{DBLP:conf/icdt/BunemanKT01}, which is to understand where data comes from, how it is collected and how it can be best used, has been assuming ever-increasing importance. One problem in data provenance attracting growing attention from both academia and industry is Auditing Data Provenance (\emph{ADP}), i.e., how to audit if a given piece of data, referred to as \textit{auditing data} \cite{DBLP:conf/sp/ShokriSSS17,DBLP:conf/kdd/SongS19,DBLP:journals/tdp/RahmanRLM18}, has been used for training a machine learning model. 


The problem of ADP distinguishes itself from related research topics such as membership inference attacks with significantly different motivations and solution priorities driven by the growing needs of data-asset-based digital economy.  Two prominent examples are (I) Privacy and (II) Incentive governance \cite{yang2020federated}. Let's examine an example for each scenario: (I) Privacy: Imagine a user’s data has been collected and used to train a machine learning model without her knowledge. To protect her data privacy, an objectively rigorous and quantifiable auditing method is necessary to substantiate her challenge in potential disputes. (II) Incentive governance: In a setting where multiple parties each contributes and trades data to collectively train models in a collaborative manner, e.g., federated learning in a decentralized variant in which no central entities are governing incentivization \cite{yang2020federated}. The incentive allocation in such a scenario would necessarily entail auditing the usage of all parties’ data to a sufficiently fine granularity to guarantee trust and fairness.
From a taxonomy point of view, there are two directions for this problem: One direction is based on model-specific techniques. One example along this direction is to directly audit the target model's training process. Techniques, such as regularization and data augmentation which ``memorize'' information about the training data set in the model, have been proposed without compromising model performance \cite{DBLP:conf/ccs/SongRS17}. Unfortunately, most real auditing settings only allow access to the output or the final parameters of the target model, rather than the training process itself. Another example is to directly design a criterion on the model output to compare training data and non-training data with a preset threshold, e.g., the prediction loss \cite{DBLP:conf/sp/NasrSH19} or the prediction confidence (e.g., class probability) \cite{DBLP:conf/ndss/Salem0HBF019}. Methods along this line suffer from limited generality due to their reliance on specific criteria.






An alternative research direction is based on a shadow training technique, which has demonstrated successful application in auditing deep learning models \cite{DBLP:conf/sp/ShokriSSS17,DBLP:conf/kdd/SongS19}. The main idea is to use multiple ``shadow models'' to imitate the behavior of the target model. As the training data for the shadow models are known, the target model can be trained using the labeled outputs of the shadow models. While shadow training technique is promising for a number of scenarios, it raises two technical challenges:


$\bullet$ {\bf  Shadow model generation}.The creation of shadow models entails two necessary requirements: (I) The knowledge of the training protocol for the target model; and (II) The generation of training data for shadow models. Requirement (I) is not always guaranteed in real applications. Requirement (II) means it is necessary to generate multiple data sets based on some heuristic rules. 


    
$\bullet$ {\bf Cumulative errors}. The final auditing results depend on multiple intermediate results of machine learning models. It may cause uncertainties in practice as error accumulates.

\begin{figure}[t]
\centering
\includegraphics[height=1.1in,width=3.2in]{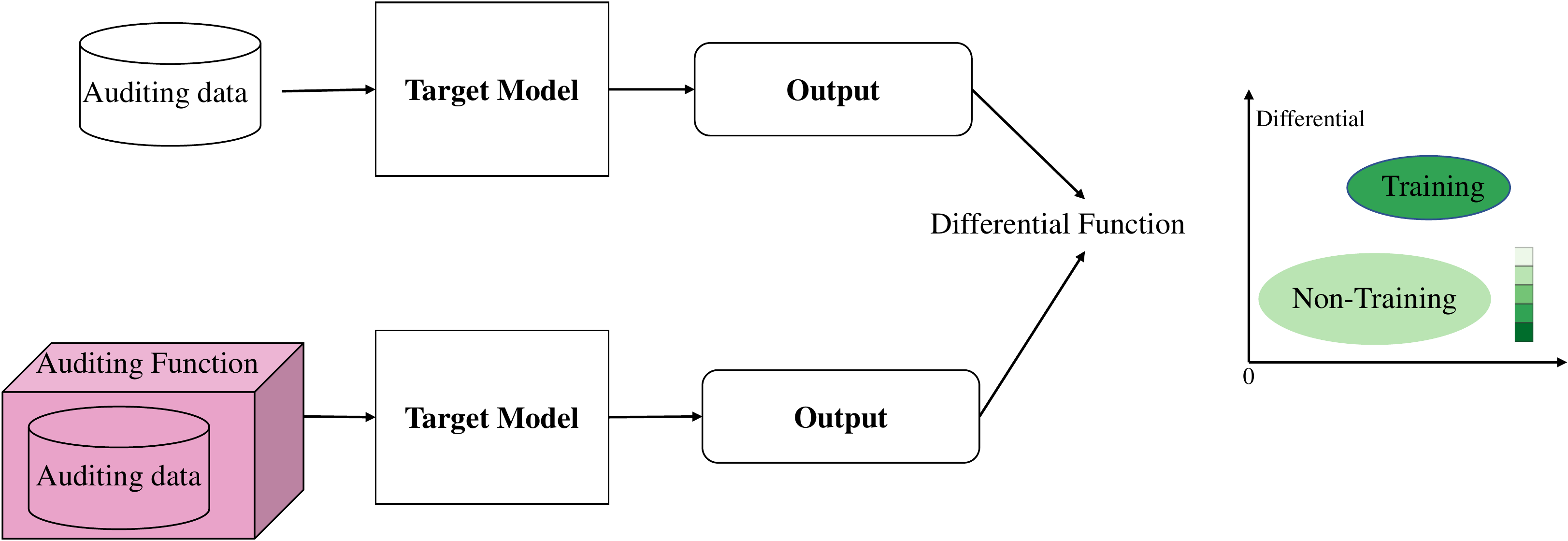}
\caption{The \textsf{DPDA} framework.}
 \label{fig:DPDA framework}
\end{figure}

In this paper, we adopt a different approach by leveraging the following observations: In general, a machine learning model tends to fit the training data well with a relatively high confidence, as, after all, the model has witnessed these data. If we apply a carefully designed function, i.e. auditing function, to transform both the training and non-training data before feeding them as input to the target model respectively, the training data side would result in a much greater difference in the target model's output between the original input data and the transformed one. This difference in the confidence of the target model's output between training and non-training data is identified as the key to the \textsf{DPDA} auditing framework we propose in this paper. As illustrated in Figure \ref{fig:DPDA framework}, the auditing framework is constituted by introducing an auditing function, which would be applied to the auditing data before feeding into the target model in one path of the comparison.


Based on this framework, we propose two implementations, an additive implementation (\textsf{DPDA}-ADD) and a multiplicative implementation (\textsf{DPDA}-MUL). \textsf{DPDA}-ADD applies additive transfer functions on input data to generate statistically significant differences, i.e., the changes in model output on training data are more significant than on non-training data. Such additive transfer functions can be easily obtained by maximizing the prediction error of the target model. \textsf{DPDA}-MUL uses a projection function to more carefully emphasize the differential between the training and non-training data, which can be learned by an alternative optimization technique to bridge the auditing model and the target model.




The proposed framework differs from the above mentioned approaches in two key aspects. First, unlike model-specific method, the proposed framework does not rely on specific model output, it can therefore be applied to a wider range of applications to multiple target models. Second, unlike shadow-based method which learns a combination of multiple shadow models to simulate target model, the proposed framework derives a data transformed directly from data through one auditing function, and the mechanism to auditing function is simpler than all these shadow-based approaches.



Another fundamentally important issue re-examined in this paper is the granularity of data that should be the subject of auditing \cite{shejwalkar2021membership}. We argue that what should be used is the notion of \emph{group-based data auditing}, where a set of data points collectively exhibit the characteristics of training or non-training data, because it reflects better the auditing needs of real applications. For example, in applications where a model has been trained with users' facial images or text sets (e.g., tweets), the real question that matters is whether a particular user's data has been used in the training, rather than whether a particular image or tweet of the user has been used. A model should be judged to have already used a user's data if a subset sufficiently characteristic of the user has indeed been used, even if some individual data points are left out in the training.

The contributions of the paper are summarized as follows:

\begin{itemize}
 
 \item We present a new framework \textsf{DPDA} for auditing data provenance with a novel notion of differential-based auditing function. Instead of auditing a specific target model in the original input data space, the \textsf{DPDA} framework distinguishes training data from non-training one by comparing a statistical differential generated by the target model between two input data spaces -- the original one and the one transformed by an auditing function. Our framework can also adapt quickly to multiple machine learning target models.
 

\item We propose two augmented auditing functions. One is additive implementation \textsf{DPDA}-ADD, the other is multiplicative implementation \textsf{DPDA}-MUL. The proposed \textsf{DPDA} targets the group-based (or user-based) \emph{ADP} problem, such that the auditor can not only infer the membership in a group of data points, but also the membership out of group, delivering a stronger solution than previous ones for the point-based problem as it easily subsumes auditing individual data points.  


\item We choose three representative benchmark datasets and one real-world application, varying from image to text data sets, to comprehensively evaluate the performance. The effectiveness of our proposed methods have been consistently demonstrated across varied experiment settings. We also study the influence of parameters in the \emph{ADP} problem and provide discussions for some important aspects of our framework for future exploration.



 \end{itemize}


\section{Related Work}
\label{sec_relatedwork}

\textbf{Membership Inference Attacks (MIAs).} The research on auditing data originated from \emph{membership inference attacks} \cite{DBLP:journals/corr/abs-2103-07853}, which is to determine if a given data record is in the model’s training data set assuming black-box model access \cite{DBLP:conf/sp/ShokriSSS17,DBLP:journals/corr/abs-1807-09173}. While MIAs has been extensively studied in many fields, e.g., computer vision \cite{DBLP:conf/ndss/Salem0HBF019,DBLP:conf/sp/NasrSH19,DBLP:conf/icml/Choquette-ChooT21}, NLP \cite{DBLP:conf/kdd/SongS19,shejwalkar2021membership,DBLP:conf/sp/PanZJY20} and recommender system \cite{10.1145/3460120.3484770}. While MIAs aims to identify training data from attackers' perspective, \emph{ADP} is motivated differently by applications in data-asset-based digital economy \cite{DBLP:conf/kdd/Pei20}. 

A major line of research is to retrain an inference model to simulate the target model, and then use the inference model to generate multiple results to make final predictions. In \cite{DBLP:journals/tdp/RahmanRLM18}, authors systematically study the impact of a sophisticated learning-based privacy attacks. When given a differentially private deep model with its associated utility, this paper discusses how much we can infer about the model’s training data. Hayes et. al. presented membership inference attacks against Generative Adversarial Networks (GANs) \cite{DBLP:journals/popets/HayesMDC19}. The idea is that, if a target model overfits the training data, training data will correspond to a higher confidence value on model output. In \cite{DBLP:conf/kdd/SongS19}, authors discussed how deep-learning-based text-generation models memorize their training data and provided a solution for text-generation models. These methods are often feasible in settings when certain conditions are satisfied such as the availability of label information or the knowledge of training protocols. However, in many real-life applications, it is difficult , if not impossible, to satisfy these conditions and hence the severe performance degradation of the auditing mechanisms. More recent work \cite{DBLP:journals/corr/abs-1802-04889} analysed the feasibility of membership inference when the model is overfitted or well-generalized and reported a study that discovered overfitting to be a sufficient but not a necessary condition for data auditing to succeed.





\textbf{Information Leakage.} With the rise of privacy concern for data, many works have been conducted to tackle the problem of information leakage of machine learning model. Information leakage can be grouped mainly into three types: data leakage, model leakage and training environment leakage. For example, in \cite{DBLP:conf/ccs/SongR20}, authors demonstrated that embeddings, in addition to encoding generic semantics, often also present a vector that would leak sensitive information about the input data. Deep learning models have been shown to have the ability of memorizing information \cite{DBLP:conf/iclr/ZhangBHRV17}. Recent work \cite{DBLP:conf/uss/Carlini0EKS19} showed that adversaries can extract training text from the output of text generation models, indicating memorization threats to user privacy. Note that this research area can be treated as a direct strategy when the training process is available to auditors.



\textbf{Differential Privacy (DP)} \cite{DBLP:conf/icalp/Dwork06} is studied to provide privacy preservation against membership-inference attack in the model inference stage. Many differentially private machine learning algorithms can be grouped according to the basic approaches they use to compute a privacy-preserving model \cite{DBLP:journals/corr/JiLE14}. Some approaches first learn a model on clean data and then use either the exponential mechanism or the Laplacian mechanism to generate a noise model \cite{DBLP:conf/webi/VaidyaSBH13,DBLP:conf/nips/ChaudhuriSS12}. Some mechanisms add noise to the target function and use the minimum/maximum of the noise function as the output model \cite{DBLP:journals/jpc/RubinsteinBHT12}. It also has been applied to various machine learning models including tree-based model \cite{DBLP:conf/icdm/JagannathanPW09}, neural networks \cite{DBLP:conf/ccs/AbadiCGMMT016,DBLP:conf/ccs/ShokriS15}, and federated learning \cite{DBLP:journals/corr/abs-1912-04977,DBLP:journals/corr/abs-1712-07557}. In this paper, we draw on the idea of \emph{differential} and apply a statistically significant differential for the \emph{ADP} problem.





\section{Problem Formulation}
\label{sec_definition}

\begin{table}[t]\footnotesize
  \centering
  \caption{Notation.}
  \label{Notation}
  \begin{tabular}{|c|c|}
    \hline
   \textbf{Notation}  & \textbf{Description}        \\
       \hline
     $\mathcal{M}$ &  Target model      \\
       \hline
     $\mathcal{D}^{T}$ &  Training data of $\mathcal{M}$      \\
       \hline
     $\mathcal{D}^{O}$ & Non-training data of $\mathcal{M}$    \\
        \hline
          $ D_i $ & A group of data instances   \\
        \hline
          $ |D_i|$ & Size of $D_i$   \\
        \hline
         $A()$ & Auditing function    \\
        \hline
         $ \Phi()$ & Differential calculation   \\
        \hline
          $   \theta   $ & Target model parameter   \\
        \hline
          $   W   $ & Auditing function parameter   \\
        \hline
  \end{tabular}
\end{table}

To best serve the auditing purpose, the granularity of data in this paper that an auditing algorithm is supposed to make judgement upon should be at the group level, where such a group is capable of capturing the characteristics of the underlying entity generating the data. More formally, we associate each entity $e$ with a distribution $\mathcal{A}_e$. A data set $D$ is denoted as $D\leftarrow A$ if $D$ is from distribution $A$. Two data sets $D_i$ and $D_j$ are said to be \emph{homomorphic under auditing} if they are both from $\mathcal{A}_e$, denoted as $D_i \cong D_j$. For example, $D_i$ and $D_j$ can be two sets of facial images of the same user $e$. We show notation in Table \ref{Notation}.  



We formulate \emph{ADP} as follows:
\begin{defn}{\textbf{Auditing Data Provenance (ADP).}}
Given (I) a set of data distributions  $\mathcal{A}=\{A_1,A_2,...,A_m\}$, (II) a set of groups of data instances $\mathcal{D}=\{D_1,D_2,...,D_n\}$, such that for each group $D_i \in \mathcal{D}$, $D_i\leftarrow A_k$ for some $1\le k\le m$ and $D_i = \{(x_j,y_j)\}_{j=1}^{{|D_i|}}$, where $x_j \in \mathbb{R}^d$ is a data instance and $y_j \in Y =\{1, 2,\ldots,c\}$ is its associated class label, and (III) a machine learning model $\mathcal{M}$, which has been trained on $\mathcal{D}^{T}=\{D_{k_1},D_{k_2},...,D_{k_t}\}$, $\mathcal{D}^{T} \subset \mathcal{D}$, and its correspondent class probability $P_j =\{p_1, p_2,\ldots,p_c\}$ for each input data instance $x_{j}$, the problem of \emph{Auditing Data Provenance (ADP)} is to find a function $f$ such that, for given any auditing data group $D_i \in \mathcal{D}$, \begin{equation}
\footnotesize
f(D_i, \mathcal{M}) =
\begin{cases} 
1,  & \mbox{if } \ \mbox{$\exists j, 1\le j\le t$, such that $D_i \cong D_{k_j}, D_{k_j} \in \mathcal{D}^{T}$.} \\
0, & \mbox{otherwise}.
\end{cases}
\end{equation}
When $f(D_i, \mathcal{M}) = 1$, we say model $\mathcal{M}$ has used data set $D_i$ for training, denoted as $D_i\in_A \mathcal{D}^{T}$. 
\end{defn}

Alternatively, we can define this problem as a ranking problem as follows: 
\begin{equation}
 \small
   \begin{aligned}
  \max_{g} \ \ & g(D_i,D_j)  \\
s.t. \ \  & D_i \in_A \mathcal{D}^{T}, D_j \notin_A \mathcal{D}^{T}   \\
\ \  & D_i,D_j \in \mathcal{D}, \ 1\leq i, j\leq n  
\end{aligned}
\label{ranking problem}
\end{equation}
where $g(\cdot,\cdot)$ represents a similarity function that calculates the difference between two data groups, e.g., Euclidean distance or Cosine distance, etc. The optimization problem is to find a similarity function to maximize the difference between data belonging to the target model $\mathcal{M}$'s training and non-training data. We also notice that it is evident that the group-based \emph{ADP} subsumes auditing individual data points when $|D_i|$ equals 1.

\begin{figure}[t]
\centering
\includegraphics[height=1.2in,width=3.3in]{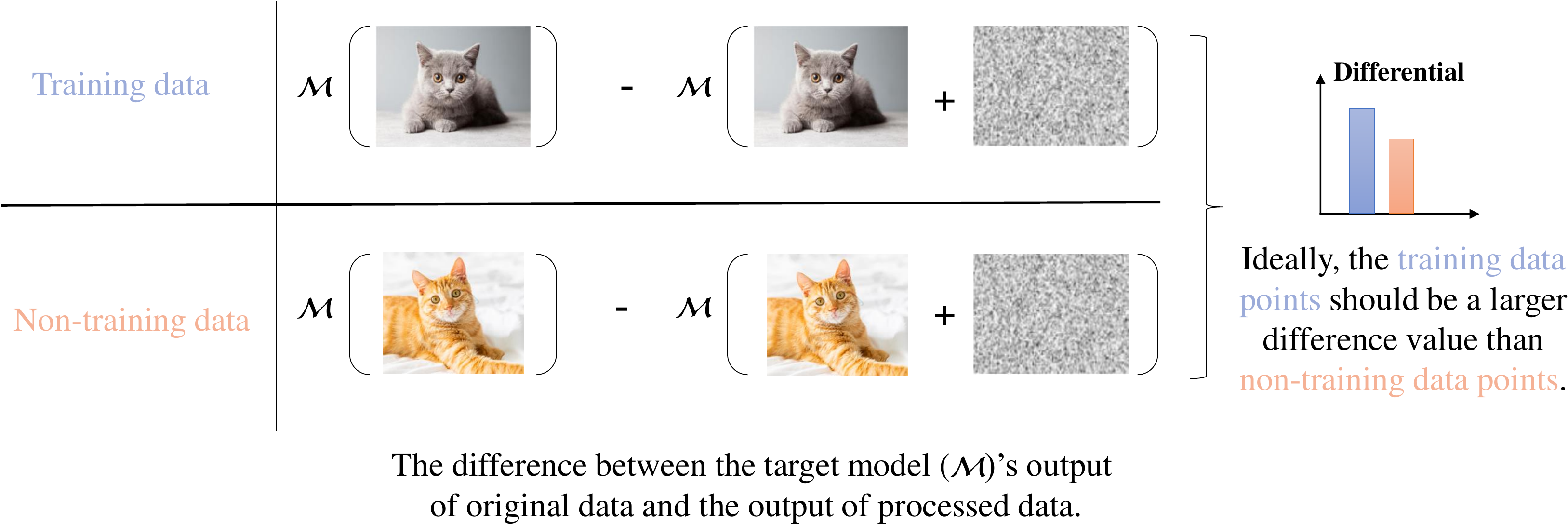}
\caption{An illustration of differential mechanism.} 
 \label{Differential mechanism}
\end{figure}

Notice that in general, we can define $\mathbb{O}$ as a distribution distance function such that $\mathbb{O}(A_{i},A_{j})$ is the distance between any two group distribution, e.g., Kullback-Leibler Divergence or Wasserstein distance. It is hard to discriminate between two data groups if the value of $\mathbb{O}$ is small, namely the two are highly similar. After all, the more similar training and non-training data groups are, the harder auditing problem becomes. In this paper, we also have provided a discussion on the issue of similarity in Section \ref{sec_dis}.

In addition, the \emph{ADP} problem is distinguished by two important conditions: the black-box condition and the white-box condition:


$\bullet$ {\bf The black-box condition (BB).} Auditors have no access to $\mathcal{M}$, i.e., no information other than model output can be accessed on model structure or model parameters. 

$\bullet$ {\bf The white-box condition (WB).} Auditors have access to all information of $\mathcal{M}$, i.e., model structure, model parameters, and model output, etc. 
\section{proposed framework}
\label{Solution}

\subsection{Design Ideas}

Our main idea is to propose a \textbf{differential mechanism} to distinguish training and non-training data by the output of the target model. The intuition is that training data directly impacts final model parameters. Carefully-designed modification on training data, if successfully transforming training data to one more similar to non-training ones, should result in greater differences in the model output, compared against the differences in model output resulted from the same modification on non-training data as, essentially, little changes have been made in terms of the nature of the input data. As illustrated in Figure \ref{Differential mechanism}, simply put, the \emph{differential mechanism} calculates the difference between the target model's output on original auditing data and the output on transformed auditing data. Ideally, training data should generate a larger difference value (represent by blue color) than non-training data.



An extreme yet straightforward case is using \textbf{differential mechanism} on instance-based machine learning models, especially those supervised learning models by storing all training instances.  \cite{DBLP:books/lib/MichalskiA84,DBLP:journals/ml/AhaKA91}. If we take a binary classification model for example, the model must output 1 for training data points, and 0 for non-training data points. It follows that, when given a piece of auditing data, if it indeed belongs to training data, the perturbation added by an auditing function would change the model output from 1 to 0, resulting in a difference of 1. On the other hand, if the auditing data belongs to non-training data, the model output remains 0 after the perturbation, resulting in a difference of 0. 



Meanwhile, there are some studies on quantifying the change of the model output by adding a perturbation, e.g., sensitivity analysis \cite{DBLP:conf/iclr/NovakBAPS18,DBLP:conf/aaai/ShuZ19,DBLP:conf/icpr/ForouzeshST20}. 
\begin{lemma} \cite{DBLP:conf/iclr/NovakBAPS18}
Consider a Gaussian perturbation $\Delta x \sim \mathcal{N}(0,\varepsilon I)$, the Frobenius norm of the class probabilities Jacobian $||\rm{J}(x)||_{\rm{F}} = \partial\mathcal{M}/\partial x^T$, we adopt the Frobenius norm $||\cdot||_{\rm{F}}$ estimates the average case sensitivity $\mathcal{M}$ around $x$:
\begin{equation*}
\begin{aligned}
\mathbb{E}_{\Delta x}[||\mathcal{M}(x)- \mathcal{M}(x+\Delta x)||_{2}^{2} ] & \cong\mathbb{E}_{\Delta x}[||\rm{J}(x) \Delta x||_{2}^{2} ] \\
&= \varepsilon ||\rm{J}(x) ||_{\rm{F}}^{2}.
\end{aligned}
\end{equation*}
\end{lemma}

\begin{lemma} \cite{DBLP:conf/icpr/ForouzeshST20}
To link the loss function to the output’s sensitivity to its input, a first order Taylor expansion can be used to show the sensitivity:
\begin{equation*}
\mathcal{M}(x + \Delta x)-\mathcal{M}(x)\cong \Delta x\cdot \nabla_x^T \mathcal{M}(x). 
\end{equation*}
\end{lemma}

\noindent Shu et. al. \cite{DBLP:conf/aaai/ShuZ19} compared network sensitivity between training and testing sets, and demonstrated the existence of difference between the two sets. They contribute to the justifiability of the \textbf{differential mechanism} underlying \emph{ADP} to consider difference between training and non-training data by the output of the target model.

\begin{figure}[t]
\centering
\subfigure[GoogLeNet on CIFAR10.]{\includegraphics[height=1.2in,width=1.6in]{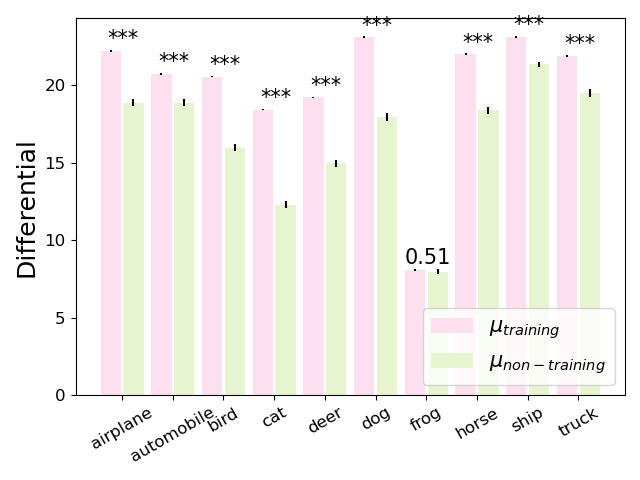}}
\hspace{.1in}
\subfigure[GoogLeNet on FansionMNIST.]{\includegraphics[height=1.2in,width=1.6in]{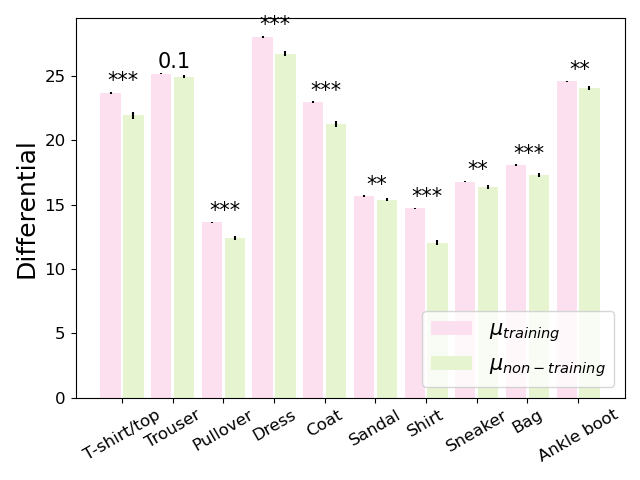}} \\
\subfigure[AlexNet on CIFAR10.]{\includegraphics[height=1.2in,width=1.6in]{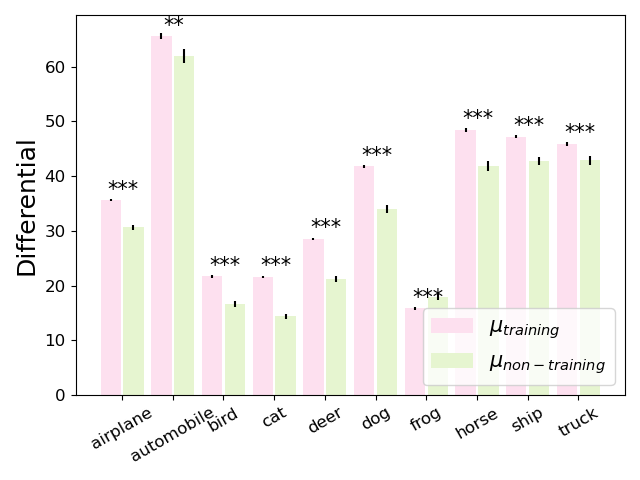}}
\hspace{.1in}
\subfigure[AlexNet on FansionMNIST.]{\includegraphics[height=1.2in,width=1.6in]{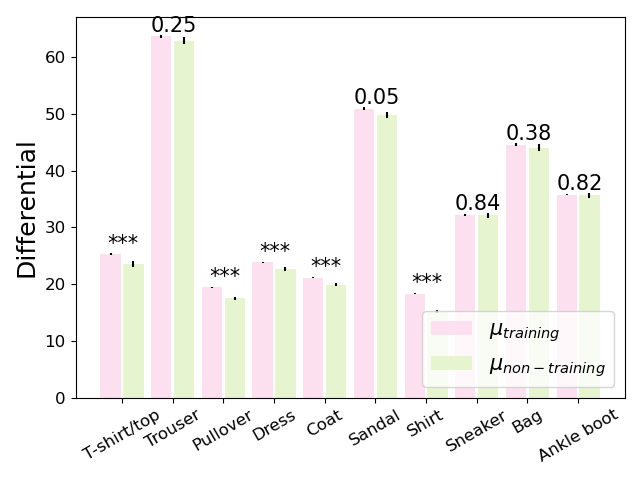}}

  \caption{The results of differential mechanism. The Y-axis is a difference value between the output of processed data point and original data point. We use average result of all training  or non-training data points per class. We check 95\% confidence intervals, and report the p-values obtained via t-tests to assess the statistical significance of differences between the average results of all training and non-training data points. $\ast: p < 0.05$; $\ast\ast: p < 0.01$; $\ast\ast\ast: p < 0.001$.} 
 \label{fig:differential mechanism}
\end{figure}

To further verify the \textbf{differential mechanism}, we examine two popular deep learning structures: GoogLeNet \cite{DBLP:conf/cvpr/SzegedyLJSRAEVR15} and AlexNet \cite{DBLP:conf/nips/KrizhevskySH12} on FashionMNIST and CIFAR-10 to observe the effectiveness of our proposed differential mechanism. The experiment is as follows: we select two popular deep learning structures: GoogLeNet and AlexNet on FashionMNIST and CIFAR-10 as observed experiments. Firstly, all auditing data (both training data and non-training data) are processed by adding Gaussian noise. Secondly, we calculate the difference value per class between the output of processed auditing data points and original auditing data points and report the average results. Note that this calculation has been done on training and non-training data separately. $p-value$ shows that there exists a gap of statistical significance between the differences of model output on training and non-training data. The results in Figure \ref{fig:differential mechanism} show that training data corresponds to a larger difference value than non-training data in most cases. This observation demonstrates the effectiveness of our proposed differential mechanism.




\subsection{\textsf{DPDA}}

Based on the differential mechanism concept, we propose a data provenance framework via differential auditing (\textsf{DPDA}), and introduce an auditing function to implement the differential mechanism as discussed in the previous section. As shown in Figure \ref{fig:DPDA framework}, \textsf{DPDA} comprises two main steps: (I) Auditing data is first processed by the auditing function; (II) The difference is calculated for the target model's output between original data and processed data. Finally, the difference is evaluated to decide whether or not the auditing data belongs to training data. 

In \textsf{DPDA}, the auditing function is chosen as a mathematical function formally defined as follows: 
\begin{defn}{\textbf{Auditing Function (AF)}:} 
Given an auditing data point $x \in \mathbb{R}^d$, the auditing function, denoted as $A()$, is defined as a bijective function such that $A(x)\in \mathbb{R}^d$ and $\forall x,x'\in \mathbb{R}^d, A(x) = A(x') \mapsto x = x'$. 
\label{define:Auditing Function}
\end{defn}
\noindent For example, if $A()$ is an additive transformation, we have $A(x) = x+\eta$, $\eta \in \mathbb{R}^d$.

\noindent{\bf Auditing Function Design.} The key to \textsf{DPDA} is to design an auditing function to embed the auditing data into a new space such that it maximizes the differential between training and non-training data, and characterizes the relation between the task of auditing data and the original task of target model.  In this paper, we propose three auditing function design as follows. It should be noted that the choices of auditing function design are not limited to these.

(1) \textbf{Offset Form.} Offset form is the most common way to do data transformation. Considering a data point $x \in \mathbb{R}^d$, an offset $z \in \mathbb{R}^d$ and a scale $\beta \in \mathbb{R}$, an auditing function in offset form is represented by $A(x) = \beta x+z$. 
    
(2) \textbf{Projection Form.} Given a data point $x \in \mathbb{R}^d$ and a matrix $V \in \mathbb{R}^{d \times d}$, the projection form is defined as $A(x)= V x$.
    
(3) \textbf{Non-linear Form.} Non-linear transfer is widely used in machine learning algorithm design, like tree-based data transfer model \cite{DBLP:conf/ijcai/ZhouF17} or activation function \cite{DBLP:journals/corr/abs-1811-03378}.


Note that while (1) and (2) have the advantage of being easy to interpret and fast to use, non-linear forms, on the other hand, are more capable to model real-world data in many cases, due to the greater complexity. 
In the following, we explore two implementations under this framework: one additive and one multiplicative. 

\section{Additive implementation}
\label{Data_dependent_sec}


We first present an additive auditing function implementation by a simple offset method as follows: 
\begin{equation}
 A(x)= x + \varepsilon \eta
   \label{section5-2}
\end{equation}
where $\varepsilon$ is a slack variable and $\eta$ is an offset.  Note that the purpose of introducing $\eta$ is to generate a larger difference between training data and non-training data. 


In general, as the target model has seen the training data, its output on training data should have higher confidence, i.e., it should be able to correctly predict data points from training data with high probability.  Consequently, for an auditing data point $x$ that is in the training data, if we can induce the processed input $A(x)$ to be \emph{misclassified}, the target model output on $x$ and $A(x)$ should then be more likely to generate a larger difference than the case if $x$ is from non-training data. We would now consider how to induce misclassification on processed data points $A(x)$.

\begin{algorithm}[t]
\renewcommand{\algorithmicrequire}{\textbf{Input:}}  
\renewcommand{\algorithmicensure}{\textbf{Output:}}  
 \caption{Additive Implementation}
   \label{data_dependnt}
\begin{algorithmic}[1]
\REQUIRE ~ $D = \{D_1,D_2,\ldots,D_e\}$ - auditing data, $\mathcal{M}$ - target model
\ENSURE ~ $D_t$ - training data
\IF {$\mathcal{M}$ is not available}
\STATE $O$ $\leftarrow$ $\mathcal{M}(D)$, \ \  $\#$ Calculate the output. 
\STATE $\mathcal{M}_r$ $\leftarrow$ train a model on $\{D, O\}$
\ENDIF
\FOR { $ i=1,...,e $}
{
\STATE $D'_i$ $\leftarrow$ $A(D_i)$, \  \      $\#$ using Eqn.(\ref{section5-2})
\STATE $S_i$ $\leftarrow$ $\Phi \left \langle \mathcal{M}(D'_i),\mathcal{M}(D_i) \right \rangle$     
}
\ENDFOR


{\IF{$\forall_j \ \  S_j > threshold$ }
 \STATE $D_j$ $\leftarrow$ training data
\ELSE{
   \STATE $D_j$ $\leftarrow$ non-training data
}
\ENDIF
}
\end{algorithmic}
\end{algorithm}


To that end, we review the adversarial example learning perspective \cite{DBLP:journals/corr/abs-1810-00069}. An adversarial example is a widely-used way to conduct an attack. Attackers alter inputs by adding small, often imperceptible, perturbations to force a learned classifier to misclassify the resultant adversarial inputs, which would still be correctly classified by a human observer \cite{DBLP:conf/iclr/KurakinGB17a,DBLP:conf/iclr/KurakinGB17}. Goodfellow et.al. \cite{DBLP:journals/corr/GoodfellowSS14} provided a strategy to use the linear view to generate adversarial inputs. Let $\theta$ be the parameters of a model, $x$ be the input to the model, $y$ be the label associated with $x$, $J(\theta)$ be the cost to train the model, $\varepsilon$ is a slack variable and the adversarial example can be defined by 
\begin{equation}
   x^{'}= x + \varepsilon \cdot sign[\nabla_{x}J(\theta)]
   \label{section5-1}
\end{equation}

Following this idea, we set $\eta$ in Eqn. (\ref{section5-2}) as the sign of the target model’s cost function gradient
\begin{equation}
  \eta = sign[\nabla_{x}J(\theta)]. 
   \label{section5-eta}
\end{equation}
This setting would maximize the loss function and result in the greatest misclassification for the auditing data processed by $A()$. 

Algorithm \ref{data_dependnt} illustrates the sketch of the additive implementation. Lines 5-13 show the case when the target model is under the white-box assumption. When the target model is under the black-box assumption, we need to build an extra machine learning model to imitate the prediction behaviors of the target model. We employ a simulation model trained by auditing data and its model output (e.g., SVM-based model). Then we use this simulation model to calculate $\eta$. The process is described in Algorithm \ref{data_dependnt} Lines 1-4. 







\noindent \textbf{Threshold determination.} In Algorithm \ref{data_dependnt} Line 9, a threshold is needed in Algorithm \ref{data_dependnt} to decide whether the auditing data belongs to training data.  As mentioned in previous discussions, we expected that training data have a larger value of differentials than non-training data, that is to say, $\{S_1,S_2,\ldots,S_e\}$ should form two groups in distributions. We adopt the following method  \cite{DBLP:journals/tkde/MuTZ17} to identify the best threshold to separate the two groups. We first generate a list $Q$ of all values in $\{S_1,S_2,\ldots,S_e\}$ in descending order. A threshold $\tau$ in this list yields two sub-lists, $Q^l$ and $Q^r$ respectively as the left sub-list and the right sub-list. The following criterion minimises the difference in standard deviations $\sigma(\cdot)$:
$\hat{\tau} =  \arg \min_\tau \  |\sigma(Q^{r}) - \sigma(Q^{l})| $
The threshold $\hat{\tau}$ is used to differentiate between training and non-training data, the former corresponding to larger values and the latter smaller ones. The details are provided in the Algorithm  \ref{Determining threshold} and an example are showed in the section \ref{Illustration on synthetic data}.








\begin{algorithm}[h]
\renewcommand{\algorithmicrequire}{\textbf{Input:}}  
\renewcommand{\algorithmicensure}{\textbf{Output:}}  
   \caption{Determining Threshold}
   \label{Determining threshold}
\begin{algorithmic}[1]
  \REQUIRE $Q$ - the list of value $s$ in $\mathcal{B}$, $m$ - size of $Q$, $\tau^{\star}$ -  initialize to a larger value. $\sigma(\cdot)$ - standard deviations calculation
  \ENSURE $t^{\star}$ - threshold
 \FOR { $ i=1,...,m $}
  {
  \STATE $Q_l$ $\leftarrow$ $Q[1:i]$
    \STATE $Q_r$ $\leftarrow$ $Q[i:m]$ \ 
    \STATE $\tau$ $\leftarrow$ $|\sigma(Q[1:i]) -\sigma(Q[i:m])|$ 
 \IF {$\tau < \tau^{\star}$}  
 {
 \STATE $t^{\star}$ $\leftarrow$ $Q[i]$
 \STATE $\tau^{\star}$ $\leftarrow$  $\tau$
 }\ENDIF
 
 } \ENDFOR
\end{algorithmic}
\end{algorithm}

\begin{algorithm}[t] 
\renewcommand{\algorithmicrequire}{\textbf{Input:}}  
\renewcommand{\algorithmicensure}{\textbf{Output:}}  
 \caption{Multiplicative Implementation}
   \label{data_mapping}
\begin{algorithmic}[1]
\REQUIRE ~ $\{D_1,D_2,\ldots,D_e\}$ - auditing data, $\mathcal{M}$ - target model, $D_c$ - initialization of training data.
\ENSURE ~ $D_t$ - training data
\REPEAT
\STATE calculate W by using Eqn. (\ref{section_7})
\STATE $S_i$ $\leftarrow$ $\{\Phi \left \langle \mathcal{M}(W D_i),\mathcal{M}(D_i) \right \rangle\}_{i=1}^{e}$ 
{\IF{$\forall_j \ \  S_j > threshold$ }
 \STATE $D_j$ $\leftarrow$ training data
\ELSE{
   \STATE $D_j$ $\leftarrow$ non-training data
}
\ENDIF
}
\UNTIL Maximum number of iterations.
\end{algorithmic}
\end{algorithm}

\section{multiplicative Implementation}
\label{mapping_method_sec}

In this section, we introduce the  multiplicative auditing functions. We have shown in Eqn. (\ref{ranking problem}) that the \emph{ADP} problem can be treated as a ranking problem. It follows that it can be turned into a differential optimization problem, and we can search for auditing function $A()$ by the following objective function:
\begin{equation}
\small
\begin{aligned}
   \max \ \ \Phi \left \langle \mathcal{M}(A(D_{t})),\mathcal{M}(D_{t}) \right \rangle -\Phi\left \langle\mathcal{M}(A(D_{o})),\mathcal{M}(D_{o})\right \rangle
\end{aligned}
\label{section_4}
\end{equation}
where $\mathcal{M}(\cdot)$ is the target model output, $A(\cdot)$ is the auditing function, $D_{t}$ is $\mathcal{M}$'s training data, $D_{o}$ is the non-training data and $\Phi(\cdot,\cdot)$ represents the differential calculation function. Since Eqn. (\ref{section_4}) would aim for the maximum difference, the training and non-training data would therefore exhibit significant gaps for the differential results.

Specifically in this work, we define a multiplicative implementation by a projection as follows:
\begin{equation}
    A(x)=Wx,
\end{equation}
where $W \in \mathbb{R}^{d \times d}$. Thus, Eqn. (\ref{section_4}) can be transformed to the following problem:
\begin{equation}
\small
\begin{aligned}
   \max \ \ \Phi \left \langle \mathcal{M}(W (D_{t})),\mathcal{M}(D_{t}) \right \rangle -\Phi\left \langle\mathcal{M}(W (D_{o})),\mathcal{M}(D_{o})\right \rangle
\end{aligned}
\label{mapping_method_eq}
\end{equation}





\noindent\textbf{Optimization.} It is important to note that Eqn. (\ref{mapping_method_eq}) requires optimization on $D_t$ and $W$ simultaneously. We employ an alternating optimization algorithm to solve it. We initialize labeled auditing data for learning $W$ by the following steps: Randomly initialize $W$, and calculate the value $\Phi$ of auditing data $D$. Then calculate a threshold (as mentioned in the section \ref{Data_dependent_sec}) to separate the value $\Phi$, and label training data $D_t$ and training data $D_o$. The optimization procedure is as follows:

(1) We consider $W$ as a variable. $D_t$ and $D_o$ are set as described above. The gradient descent technique is then applied to efficiently solve Eqn. (\ref{mapping_method_eq}).

(2) After $W$ is obtained, we calculate the value $\Phi$ of $D$.

(3) Calculate the threshold (Algorithm \ref{Determining threshold}) to separate the value $\Phi$ of auditing dataset $D$. The larger ones are set as $D_t$, the others as $D_o$.


(4) Use the newly updated $D_t$ and $D_o$ to calculate $W$. 

\noindent The procedure stops when the terminating condition is satisfied, i.e., a predetermined maximum number of iterations. 

Note that we can apply the gradient descent algorithm to efficiently update $W$ as follows:
\begin{equation}
\begin{aligned}
   W' = W -  \epsilon \frac{\partial J(W)}{\partial W}
\end{aligned}
\label{section_7}
\end{equation}
where
\begin{equation}
\begin{aligned}
   \frac{\partial J(W)}{\partial W} = & [\mathcal{M}(W D_t)-\mathcal{M}(D_t)] \mathcal{M}^{'}(W D_t) D_t \\
   & -[\mathcal{M}(W D_o)-\mathcal{M}(D_o)] \mathcal{M}^{'}(W D_o) D_o
\end{aligned}
\label{section_8}
\end{equation}
The sketch of the process is described in Algorithm \ref{data_mapping}. Note that when $\mathcal{M}^{'}$ is not available under the black box assumption, a simulation model can be employed. 






\section{Experiment}
\label{EXP}


%




\begin{table*}[t]\footnotesize
  \centering
  \caption{Results of different target models on different data sets.}
  \label{tbl_bench1}
  \begin{tabular}{c|c|c|c|c|c|c|c}
   \hline
   &  & \multicolumn{2}{c|}{MNIST} & \multicolumn{2}{c|}{20 Newsgroups}&\multicolumn{2}{c}{CIFAR10} \\
    \hline
   &   Algorithm &  F-measure & AUC & F-measure & AUC& F-measure & AUC\\
    \hline
\multirow{5}{*}{SVM-based model}   &  CC &   0.412 $\pm$ 0.02 & 0.513 $\pm$ 0.01 & 0.401 $\pm$ 0.02 &  0.535 $\pm$ 0.01 & 0.534 $\pm$ 0.02 & 0.535 $\pm$ 0.01  \\
  &  SLT &   0.612 $\pm$ 0.05 & 0.805 $\pm$ 0.04 & 0.715 $\pm$ 0.03 &  0.711 $\pm$ 0.05 & 0.704 $\pm$ 0.05 & 0.750 $\pm$ 0.03  \\
  &   RN &  0.536 $\pm$ 0.03 & 0.550 $\pm$ 0.02 & 0.433 $\pm$ 0.02 &  0.565 $\pm$ 0.02 & 0.586 $\pm$ 0.03 & 0.526 $\pm$ 0.02  \\
  &   ADD &   0.695 $\pm$ 0.04 & 0.822 $\pm$ 0.06 & 0.660$\pm$ 0.02 & 0.735 $\pm$ 0.02 & 0.700 $\pm$ 0.02 & 0.803 $\pm$ 0.04   \\
&       MUL  &0.719 $\pm$ 0.06 & 0.821 $\pm$ 0.05 & 0.723 $\pm$ 0.02 &  0.750 $\pm$ 0.02 & 0.726 $\pm$ 0.02 &0.801 $\pm$ 0.04\\
      \hline
\multirow{5}{*}{Tree-based model}   &  CC &   0.423 $\pm$ 0.02 & 0.533 $\pm$ 0.01 & 0.339 $\pm$ 0.01 &  0.554 $\pm$ 0.02 & 0.540 $\pm$ 0.04 & 0.573 $\pm$ 0.04  \\
  &  SLT &   0.696 $\pm$ 0.03 & 0.711 $\pm$ 0.04 & 0.655 $\pm$ 0.02 &  0.684 $\pm$ 0.02 & 0.744 $\pm$ 0.02 & 0.753 $\pm$ 0.05  \\
  &   RN &  0.671 $\pm$ 0.05 & 0.654 $\pm$ 0.06 & 0.632 $\pm$ 0.06 &  0.652 $\pm$ 0.06 & 0.651 $\pm$ 0.07 & 0.673 $\pm$ 0.04  \\
  &   ADD &   0.675 $\pm$ 0.03 & 0.652 $\pm$ 0.05 & 0.653 $\pm$ 0.03 & 0.654 $\pm$ 0.02 & 0.675 $\pm$ 0.02 & 0.654 $\pm$ 0.05   \\
&       MUL  &0.695 $\pm$ 0.01 & 0.712 $\pm$ 0.04 & 0.666 $\pm$ 0.02 &  0.704 $\pm$ 0.03 & 0.760 $\pm$ 0.03 &0.772 $\pm$ 0.02\\
    \hline
\multirow{5}{*}{NN-based model} &    CC &   0.493 $\pm$ 0.01 & 0.565 $\pm$ 0.03 & 0.432 $\pm$ 0.02 &  0.515 $\pm$ 0.02 & 0.478 $\pm$ 0.02 & 0.523 $\pm$ 0.01  \\
  &   SLT &   0.743 $\pm$ 0.03 & 0.811 $\pm$ 0.02 & 0.693 $\pm$ 0.01 &  0.701 $\pm$ 0.01 & 0.721 $\pm$ 0.02 & 0.798 $\pm$ 0.03  \\
   &  RN &  0.521 $\pm$ 0.01 & 0.554 $\pm$ 0.02 & 0.442 $\pm$ 0.01 &  0.542 $\pm$ 0.01 & 0.571 $\pm$ 0.01 & 0.563 $\pm$ 0.01  \\
  &   ADD &   0.739 $\pm$ 0.04 & 0.792 $\pm$ 0.02 & 0.703 $\pm$ 0.03 & 0.724 $\pm$ 0.02 & 0.726 $\pm$ 0.04 & 0.803 $\pm$ 0.03   \\
    &   MUL  &0.782 $\pm$ 0.03 & 0.802 $\pm$ 0.01 & 0.723 $\pm$ 0.02 &  0.751 $\pm$ 0.02 & 0.719 $\pm$ 0.01 &0.810 $\pm$ 0.04 \\
    \hline   
    \multicolumn{2}{c|}{ADD and MUL have \#wins/\#draws/\#losses}& \multicolumn{2}{c|}{3/3/0 }& \multicolumn{2}{c|}{6/0/0 }&\multicolumn{2}{c}{5/1/0}\\
       \hline
  \end{tabular}
\end{table*}

\subsection{Experiment Setup}

{\bf Data Sets. } We use 
four datasets to compare the performance of all methods: {\bf MNIST}, {\bf 20 Newsgroups}, {\bf CIFAR-10} and {\bf VGGFace}.

\noindent {\bf Competing Algorithms.} A brief description of each of the methods used in the experiment is given as follows

(1) {\bfseries Confidence Criteria (CC)}: CC means we directly design a criterion on the model output (e.g., the perdition class probability). The criterion is like a preset threshold, the data with higher probability is treated as training data.


(2) {\bfseries Shadow Learning Technique (SLT)} \cite{DBLP:conf/sp/ShokriSSS17}: SLT introduces multiple shadow models and an attack model to address the data auditing problem. The shadow model is to recognize differences in the target model’s predictions on the inputs that it has trained on versus the inputs that it has not trained on. The attack model is treated as a classifier to distinguish the output of shadow model.

(3) {\bfseries \textsf{DPDA}-RN} (RN for short): \textsf{DPDA} with the additive implementation but by setting random values.

(4) {\bfseries \textsf{DPDA}-ADD} (ADD for short): \textsf{DPDA} with 
the additive implementation. 


(5) {\bfseries \textsf{DPDA}-MUL} (MUL for short): \textsf{DPDA} with the multiplicative implementation.


\noindent{\bf Experiment Settings.} All experiments are implemented in Python on Intel Core CPU machine with 128 GB memory and NVIDIA RTX 3090 GPU. The following implementations are used: In CC, the confidence of one instance is set by the largest value of its estimated label probability. In SLT, the codes are developed based on the original paper\footnote{https://github.com/spring-epfl/mia}. We employ 50 shadow model and one attack model which is SVM\footnote{https://scikit-learn.org/stable/modules/svm.html} with RBF kernel and other parameters are set by default values. In DPDA-RN, random values are set by Gaussian noise. The parameter $\varepsilon$ in \textsf{DPDA}-ADD is set by $[e^{-8},e^8]$. In particular, we can use data similarity as a measure to guide the setup: the higher the data similarity, the smaller the value $\varepsilon$. In \textsf{DPDA}-MUL, $W$ is initialized to a semi-positive definite matrix.




\noindent{\bf Evaluation Metrics.} We use \textbf{F-measure} and \textbf{AUC}\footnote{https://en.wikipedia.org/wiki/Receiver\_operating\_characteristic} to measure performance. As mentioned in the definition of \emph{ADP}, auditing data are in the form of groups. We conduct both AUC and F-measure on group-based dataset as follows: First, each auditing data is fed as input for prediction, and we calculate the average results of each group. Then, the AUC or F-measure results are calculated in those group results. Note that each group contains the same type of data, i.e., either all training data or all non-training data.

\begin{figure}[t]
  \centering
\subfigure[Auditing data and target model.]{\includegraphics[height=1.3in,width=1.6in]{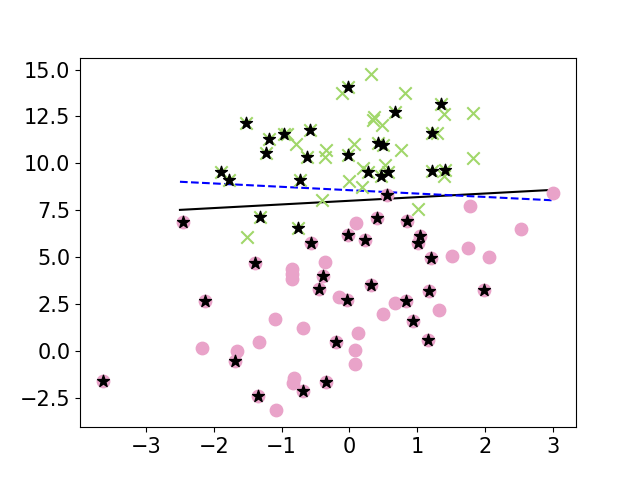}}
 \hspace{.2in}
\subfigure[Results under WB condition.]{\includegraphics[height=1.3in,width=1.6in]{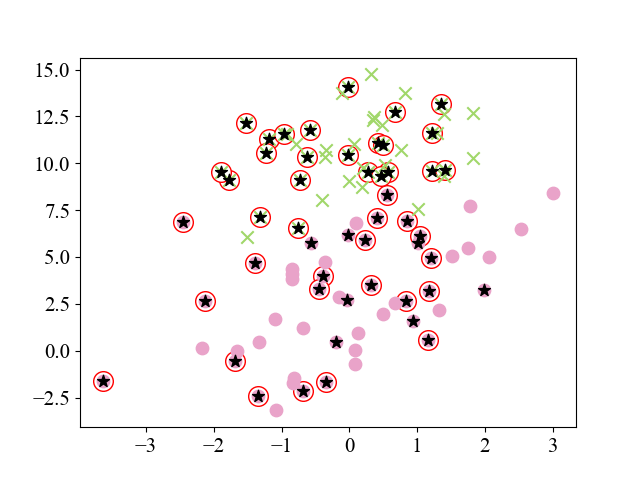}} \\
\subfigure[Results under BB condition.]{\includegraphics[height=1.3in,width=1.6in]{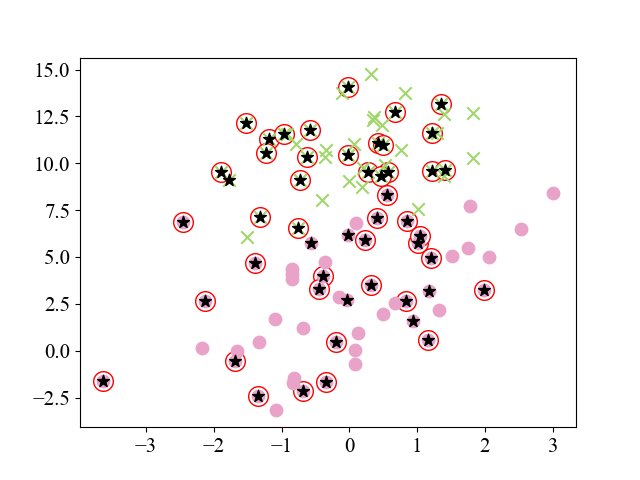}}
\hspace{.2in}
\subfigure[Threshold setting.]{\includegraphics[height=1.2in,width=1.4in]{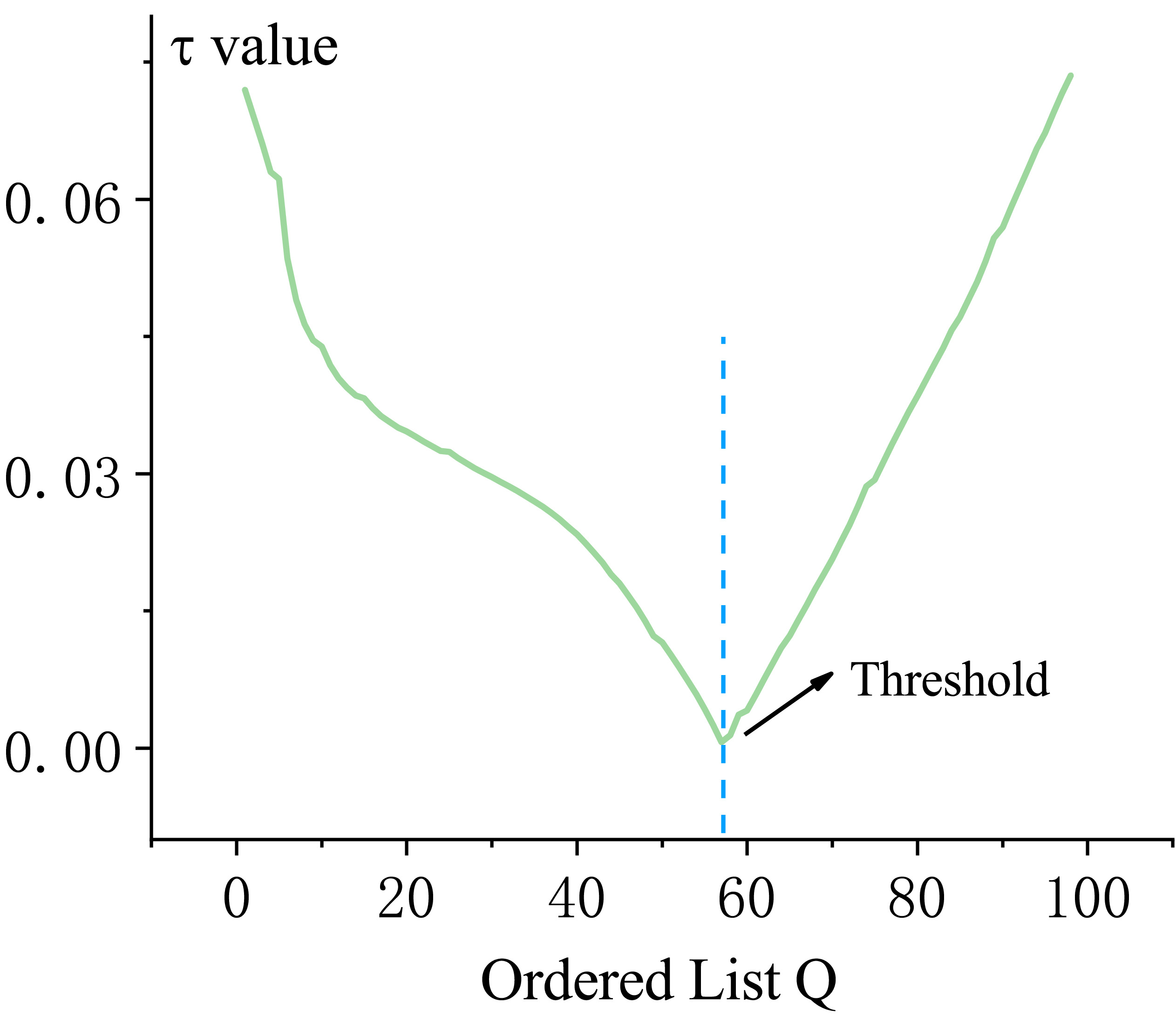}}
  \caption{An illustration on SVM-based model.} 
\label{examplesvm}
\end{figure}

\subsection{Results on synthetic data} 
\label{Illustration on synthetic data}


We take SVM-based, Tree-based and NN-based model as the target model and train them on a two-dimensional synthetic dataset with 100 data points. Figure \ref{examplesvm} shows SVM-based model results. We indicate two kinds of data in green and red respectively, and mark training data by ``black star''. The auditing problem is to identify the ``black star'' training data given all data points as the input auditing data. 




Figure \ref{examplesvm}(a) shows the target model by a solid black line and a simulation model by blue dotted lines. The effectiveness of the proposed method is demonstrated by the auditing results as labeled with red circles in Figure \ref{examplesvm}, in which (b) shows the white-box condition and (c) shows the black-box condition. In addition, we show the threshold setting in Figure \ref{examplesvm}(d). The 100 data points in the synthetic data set are divided into training and non-training groups with the ratio of 1:1. Figure \ref{examplesvm}(d) shows an example of the distribution for $\tau$ ($=|\sigma(Q^{r}) - \sigma(Q^{l})|$) curve. Note that the lowest point provides a clear guide to separate the auditing data into the two parts of training and non-training, and it is close to the optimal value of 50.


\subsection{Results on benchmark datasets}
\label{results}




\noindent\textbf{Setting.} Each dataset is used to simulate the following auditing environment. We first randomly select two classes, and instances of these two classes are selected as an auditing data set, which are then randomly divided into training data  and non-training data with the ratio between them being roughly 1:1. The training set is then used to train a target model to be audited. Subsequently the auditing data are grouped into multiple subsets, each containing the same type of data, i.e., all training data or all non-training data. All experiments are under the BB assumption. Target models are SVM-based models, Tree-based models and NN-based models respectively: SVM-based models are set by a least squares SVM classifier; Tree-based models use random forest with completely random trees; NN-based models are set by two fully connected layers and a SoftMax layer. Parameters of target models are set by default according to their official code package. We run 30 independent experiments with different simulations on each dataset.


\noindent\textbf{Summary.} Table \ref{tbl_bench1} provides more comprehensive results on SVM-based, Tree-based and NN-based target model. Our proposed \textsf{DPDA} models, both \textsf{DPDA}-ADD and \textsf{DPDA}-MUL, have produced higher AUC performance in all data sets than all other methods. The closest contender SLT, which is based on shadow model technique, is weaker than \textsf{DPDA}. \textsf{DPDA}-RN is based on random perturbation and its performance falls behind \textsf{DPDA} in all data sets. The performance of CC ranks at the bottom. An analysis is provided below:






$\bullet$ CC performs worse than others in all data sets. This shows that efforts to directly use the prediction probability are unsuccessful for the \emph{ADP} problem. There are a couple of reasons for this, e.g., the distribution of the model output could be dense and therefore makes the separation of training and non-training data difficult. It is thus concluded that CC is not a good choice for this task.
    
$\bullet$ SLT requires training multiple shadow models and an attack classification model. It is important to note that its performance highly dependents on shadow model results. Unsatisfactory results of shadow model will severely limit the attack model's ability for classification. In addition, SLT needs label information for training shadow models, which is often hard to obtain in real applications. Nevertheless, the experimental results show that it still performs worse than the \textsf{DPDA} framework in two out of three data sets.
    
$\bullet$ RN presents worse performance than other \textsf{DPDA} models except on the tree-based model. The under-performance of auditing functions of random values drives home the effectiveness of the two augmentations we proposed for auditing functions. 
    
$\bullet$ Both ADD and MUL are demonstrated to be competitive methods for the \emph{ADP} problem. While MUL achieves a higher performance, ADD excels with its lower computational cost in an extensive parameter search and easier implementation. The choice between them in real applications should be a result of comprehensive consideration on case-dependent factors. 

\subsection{Results on a real-world application}
\label{parameterreal-worlddata}
Gender estimation is an important and challenging task in many real-world applications. Over the past few years, most methods used deep learning models to estimate gender achieved respectable results \cite{DBLP:journals/access/ZhangGGSYHZL17,DBLP:conf/bigdataconf/SmithC18,DBLP:journals/tifs/EidingerEH14}. In this section, we evaluate \textsf{DPDA} and contenders on auditing a gender estimation model. The gender estimation model is set by a famous computer vision machine learning model, ResNet18 \cite{DBLP:conf/cvpr/HeZRS16}, including 18 layers deep neural network. And it is trained on VGG-face dataset \cite{DBLP:conf/bmvc/ParkhiVZ15}. The aims of this section are to examine the ability of \textsf{DPDA} to (i) adapt to a real-world group-based auditing problem; and (ii) achieve a good performance.

\begin{figure}[t]
\centering
\includegraphics[height=1.5in,width=3.1in]{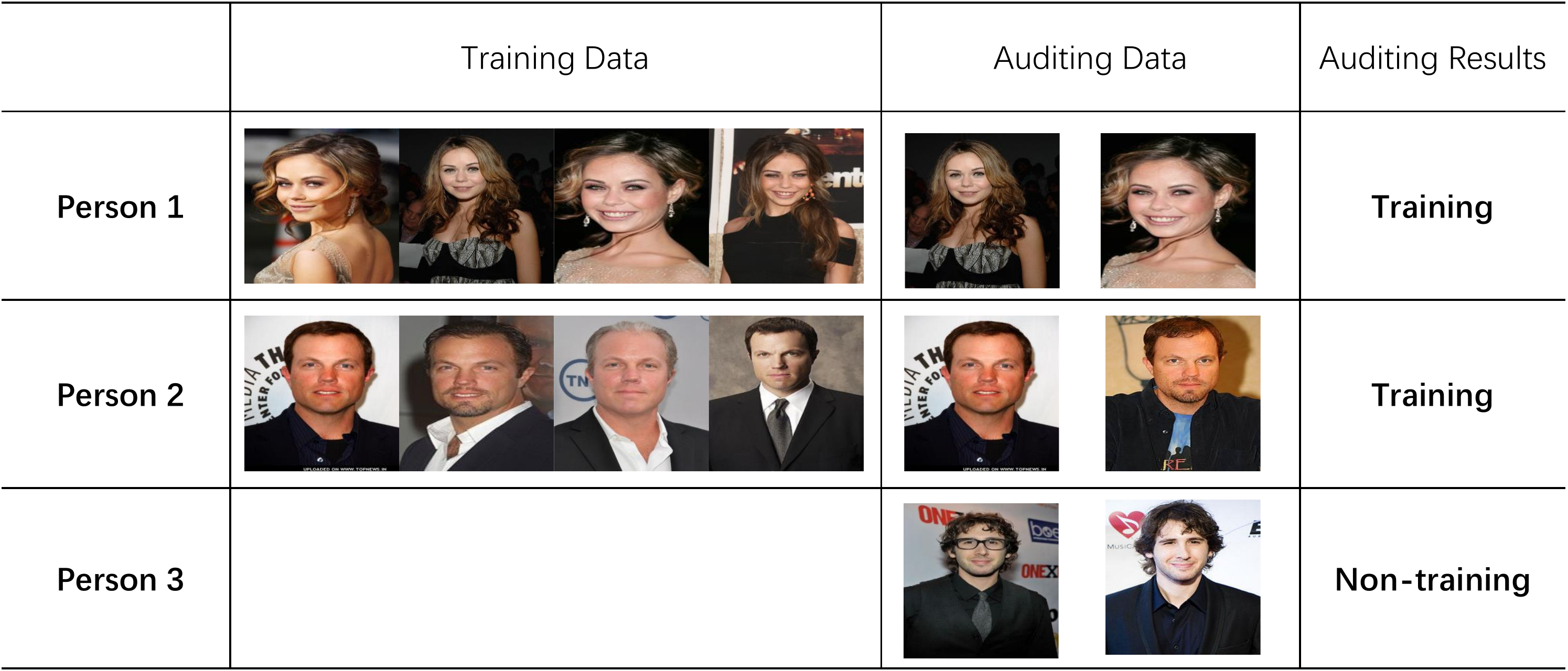}
  \caption{The illustration of the gender estimation dataset.} 
 \label{realdataill}
\end{figure}
\noindent\textbf{Setting.} In the experiment, we used 20 people with 20k images as training data to train a gender estimation classifier. The \emph{auditing data} consist of these 20 people's images which include some training images and some other images which haven't been used to train. In the auditing data, images of each people are naturally regarded as a group $D_i$. As an example shown in Figure \ref{realdataill}, person 1 and person 2 have been used to train the target model, their images are shown in Figure \ref{realdataill}. In the auditing data, the images from person 1 and person 2 are both treated as training, even some images haven't been used. Due to the data of person 3 without participating training target model, the images of person 3 are treated as non-training. This is a real application under the group-based assumption. All experiments are under the black-box assumption. Parameters of target models are set by default according to their official code package.



\begin{figure}[t]
  \centering
\subfigure[F-measure.]{
\includegraphics[height=1.1in,width=1.3in]{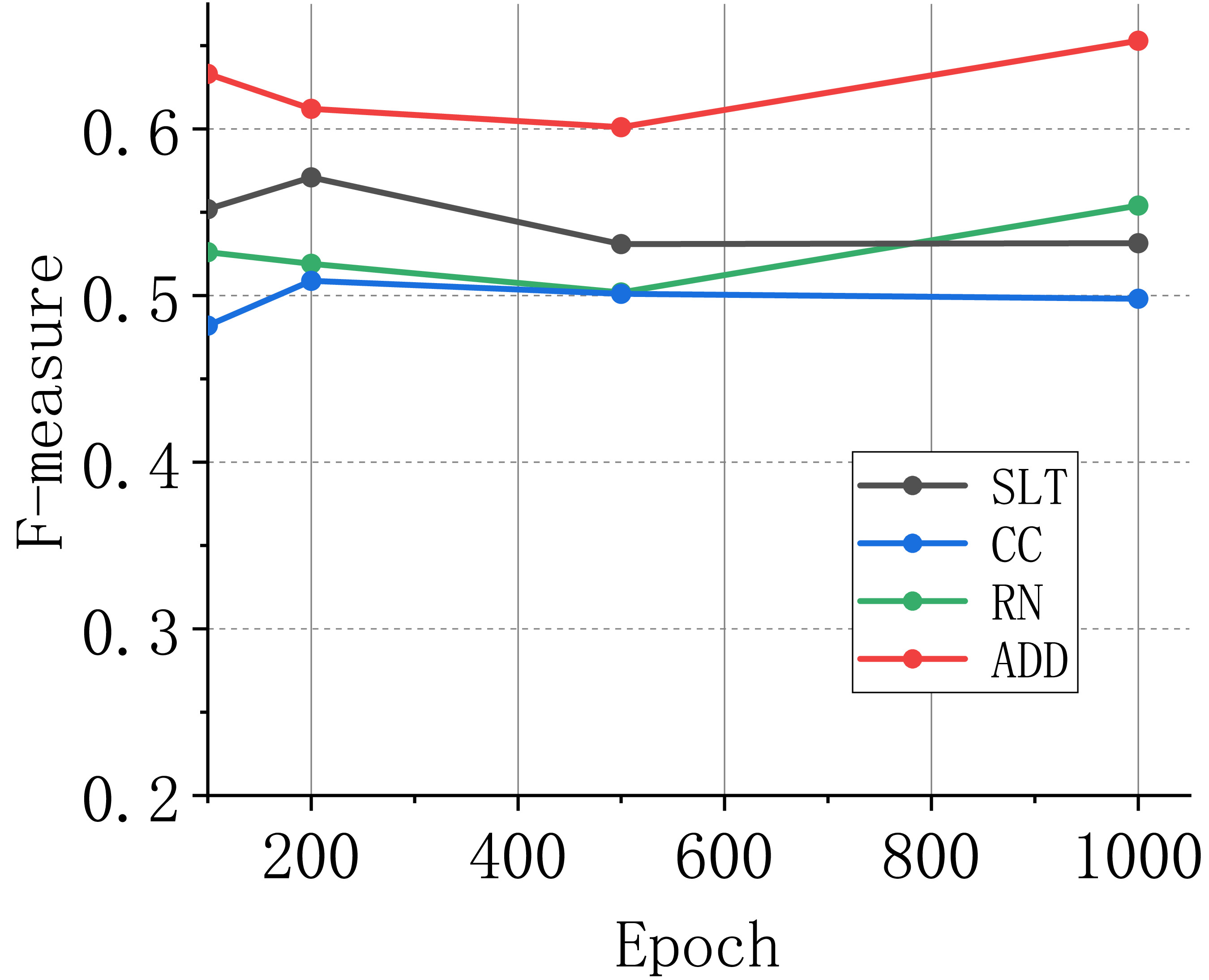}}
\hspace{.1in}
\subfigure[AUC.]{
\includegraphics[height=1.1in,width=1.3in]{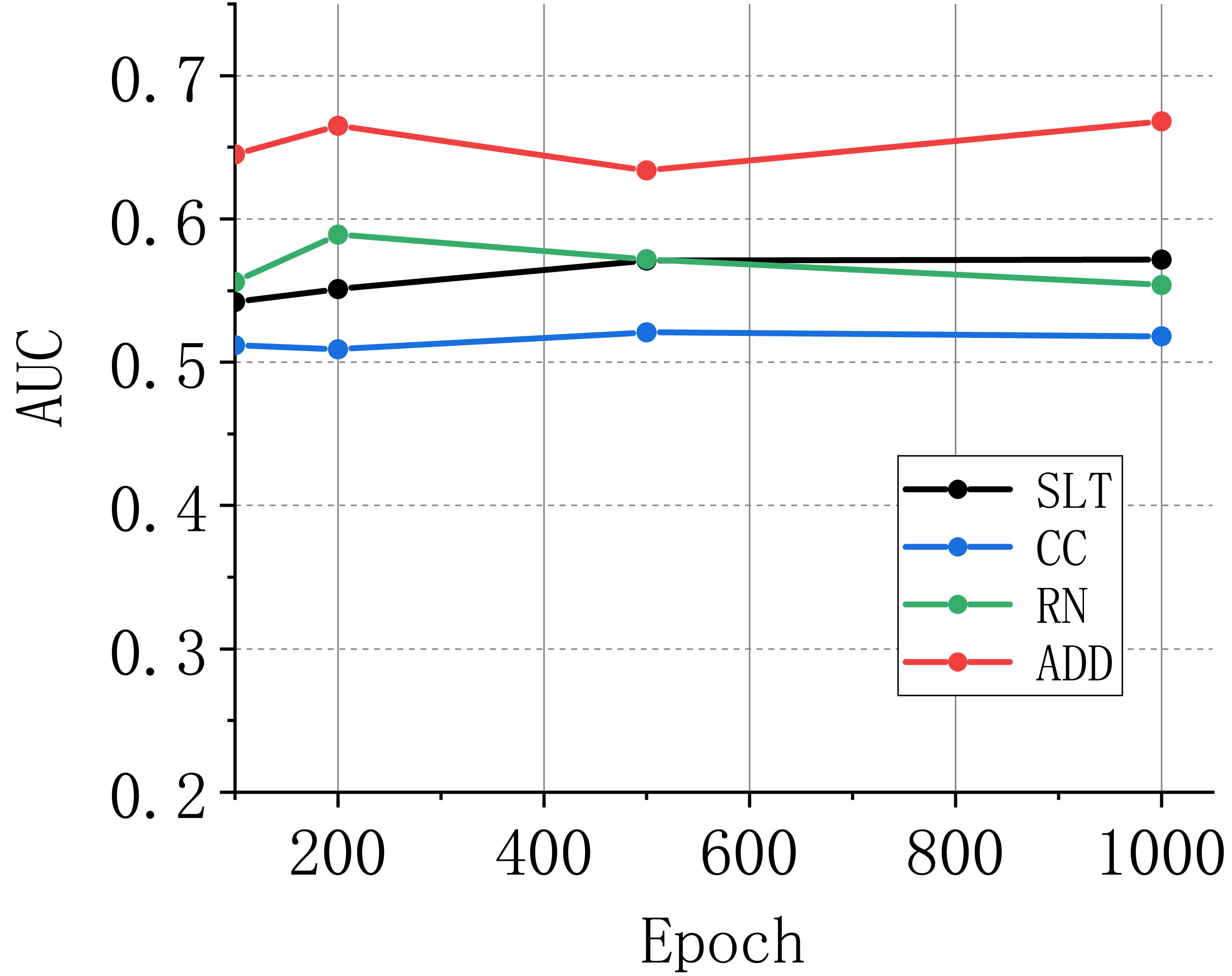}}
\caption{The results on auditing a gender classifier.}
 \label{fig:realdata}
\end{figure}

\noindent\textbf{Summary.} Figure \ref{fig:realdata} shows the auditing results over different epochs. In the different epochs, target model ResNet18 has different performance, namely accuracy is [0.64,0.78,0.79,0.82] on [100,200,500,1000]. Firstly, because the proposed method \textsf{DPDA} takes into account the global auditing results in a group, avoiding the effect of outliers, ADD performs better results than all three methods under the group-based auditing assumption. Secondly, SLT is a point-based auditing algorithm which requires to have multiple shadow models in order to simulate the target model. Despite this advantage, it still performed worse than ADD. Meanwhile, the other two baselines also perform worse than ADD. Overall, the experimental results here demonstrate the reasonableness and feasibility of this method on the group-based auditing problem.




\subsection{Parameter analysis}
\label{parameter}

\begin{figure}[t]
  \centering
  \subfigure[Different sizes of classes]{
  \includegraphics[height=1.1in,width=1.05in]{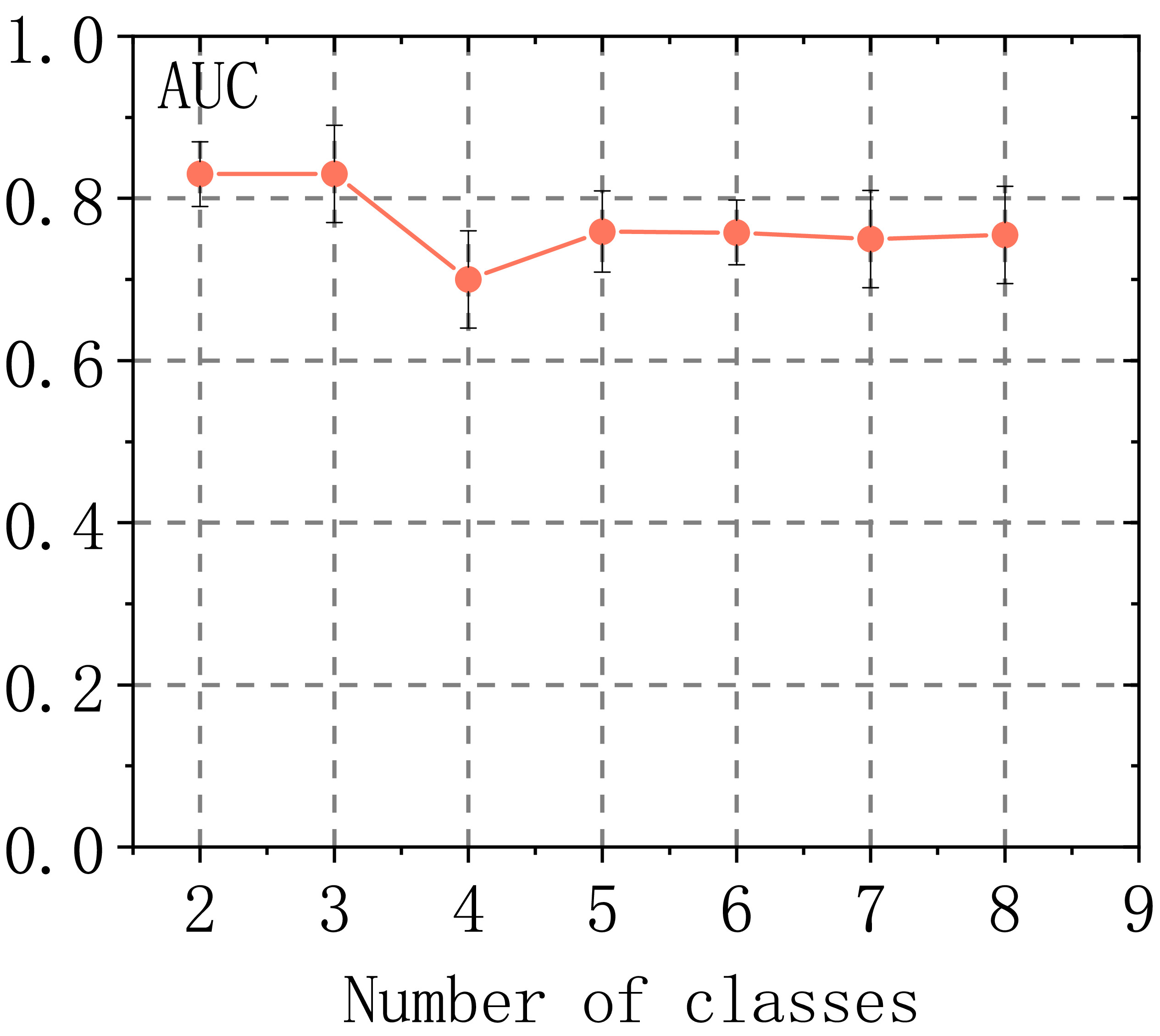}}
  \subfigure[Different sizes of data.]{
   \includegraphics[height=1in,width=1.05in]{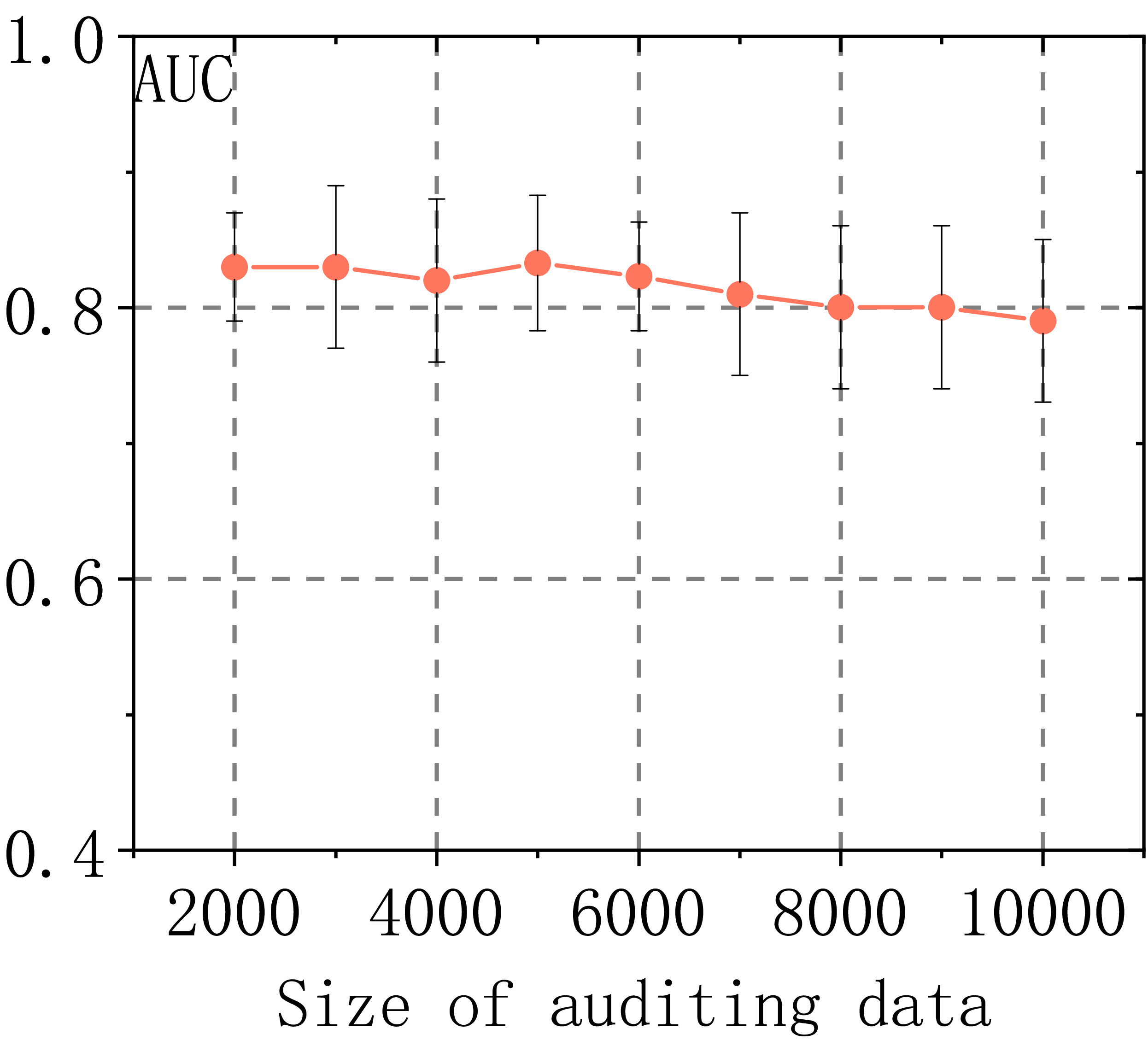}}
  \subfigure[Different numbers of features of auditing data.]{
  \includegraphics[height=1in,width=1.05in]{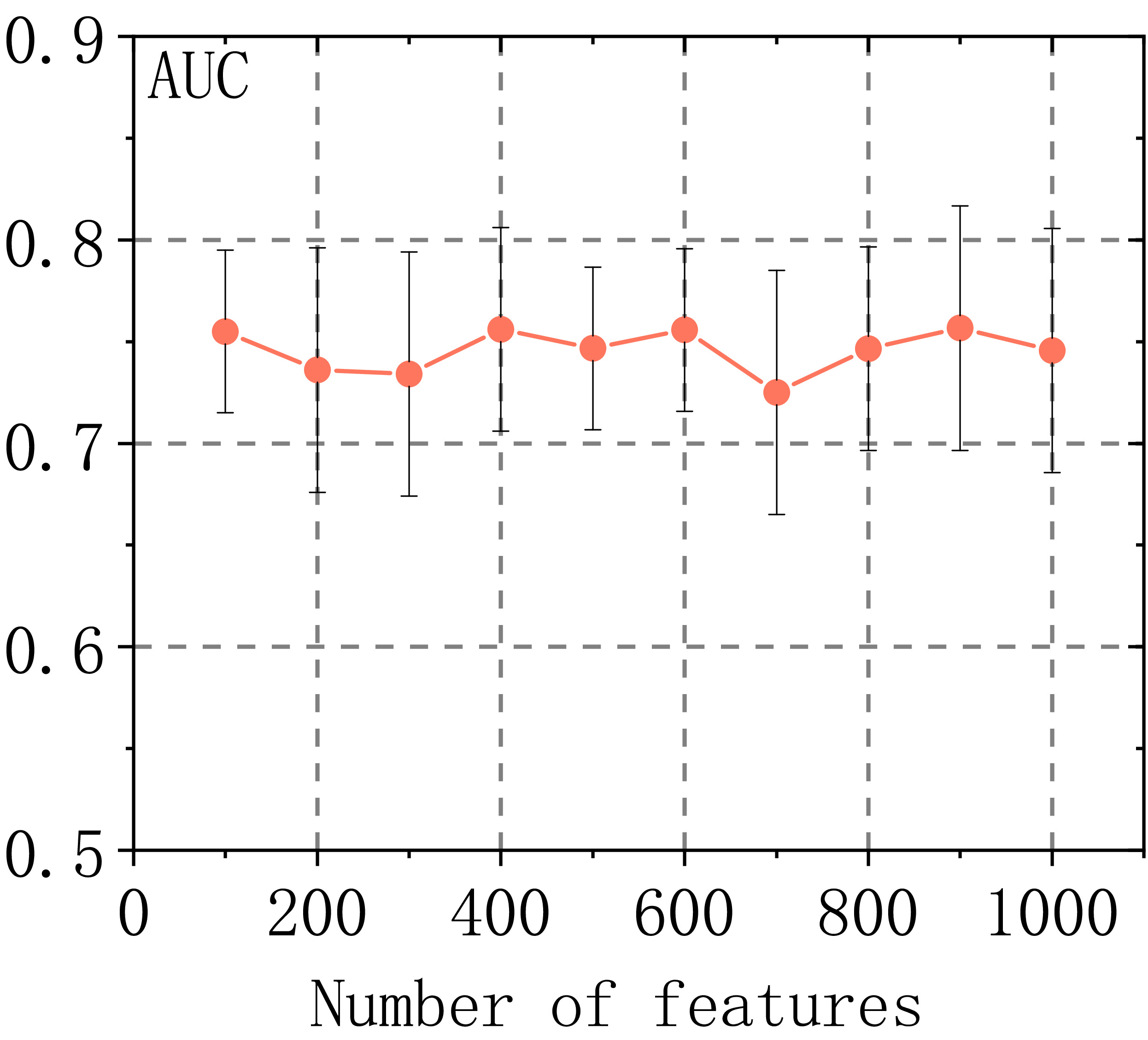}}
  \caption{Parameter analysis result.}
 \label{fig:real}
\end{figure}

\begin{table*}[t]\footnotesize
\centering
\caption{The relation between data similarity and differential mechanism. (Data similarity is calculated by average pair-wise Euclidean distance between training and non-training data, i.e., the smaller the value, the higher the similarity. The meaning of p-value is the same as Figure \ref{fig:differential mechanism}.)}
\label{Tab:relation of ds and dm}
\begin{tabular}{c|c|c|c|c|c|c|c|c|c|c}
\hline
  & `car'&	 `cat'	& `truck'&	 `dog'	 &`horse'&	`plane'	 &`ship'&	 `deer'	& `bird'&	 `frog'
   \\
      \hline
\textbf{Data similarity}  & 65.05&	63.73&	62.44&	61.86&	60.99&	58.83&	57.207	&55.78&	54.87&	 \textbf{54.21} 
 \\
        \hline
\textbf{Accuracy}  &   63\% &	 54\% & 	 69\%&	52\% &	 69\%	&  65\% & 	 70\%&	 72\% 	& 75\% 	 & \textbf{80\%} 
\\
        \hline
\textbf{p-value}  &  $\ast\ast\ast$&	$\ast\ast\ast$&	$\ast\ast\ast$&	$\ast\ast\ast$&	$\ast\ast\ast$&	$\ast\ast\ast$ &	$\ast\ast\ast$&	$\ast\ast\ast$&	$\ast\ast\ast$&	\textbf{0.507} \\
        \hline
  \end{tabular}
\end{table*}

We present a study of parameters in \textsf{DPDA}, i.e., the number of classes in auditing data set, the size of auditing data set and the number of features of auditing data. We evaluate them one at a time on varied settings with other parameters fixed. 

Figure \ref{fig:real}(a) describes the number of different sizes of classes in auditing data set and Figure \ref{fig:real}(b) describes the results of the sizes of auditing data set. We show the results of \textsf{MUL} on MNIST data set. Note that similar results are also observed under the \textsf{ADD} implementation. There are a downward trend for performance as the number of classes and the size of data increase. That said, the denser data distribution may yield a higher degree of data similarity, and make it significantly more difficult to audit data. Figure \ref{fig:real}(c) shows the number of features of auditing data. We test varied sizes of the feature vector on the 20 Newsgroups data set. Results show that the different sizes of feature vector have a relatively small impact on performance.

\section{Discussion}
\label{sec_dis}




\begin{figure}[t]
  \centering
  \subfigure[]{
  \includegraphics[height=1.2in,width=1.7in]{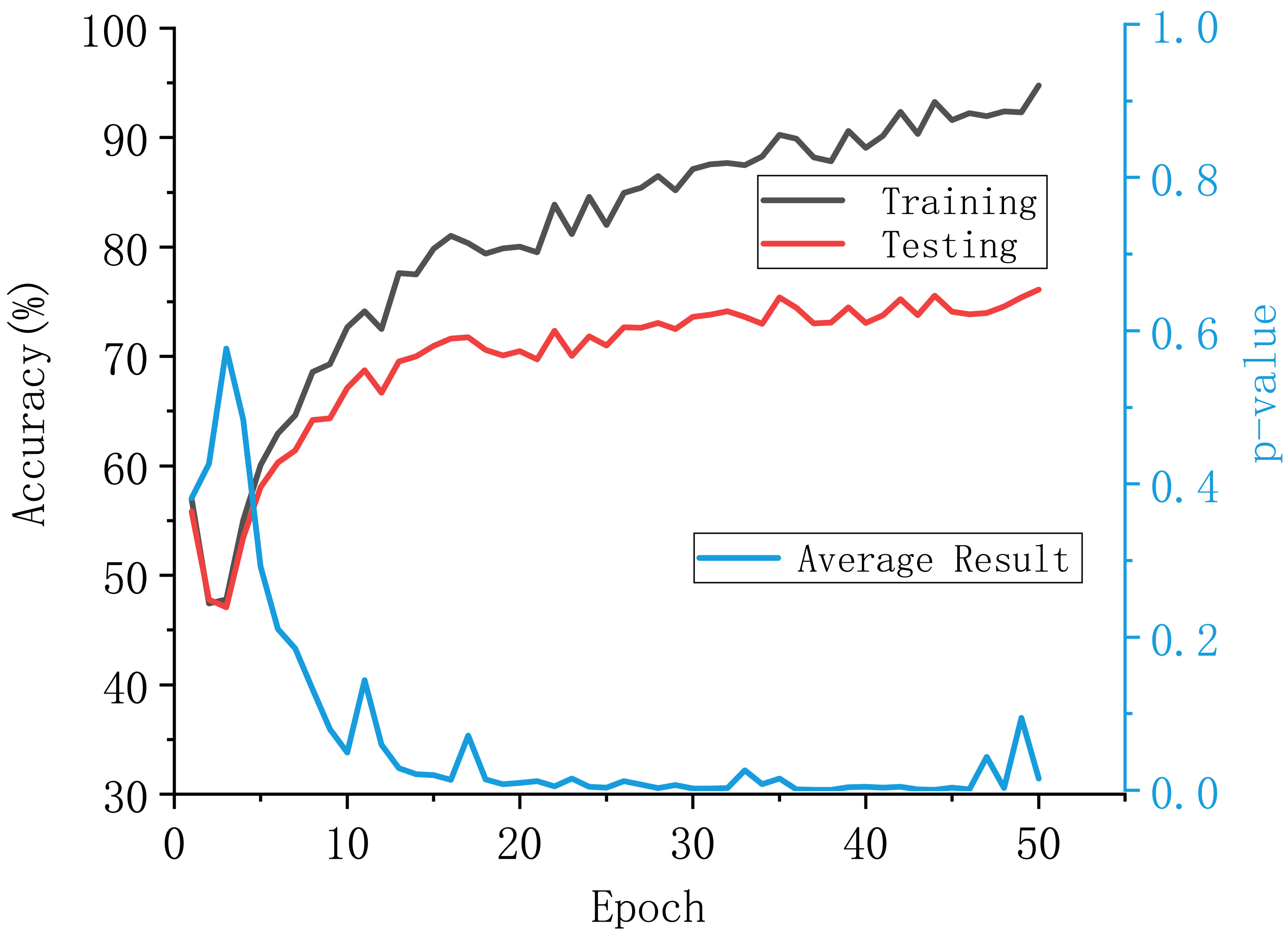}}
  \hspace{.1in}
  \subfigure[]{\includegraphics[height=1.2in,width=1.4in]{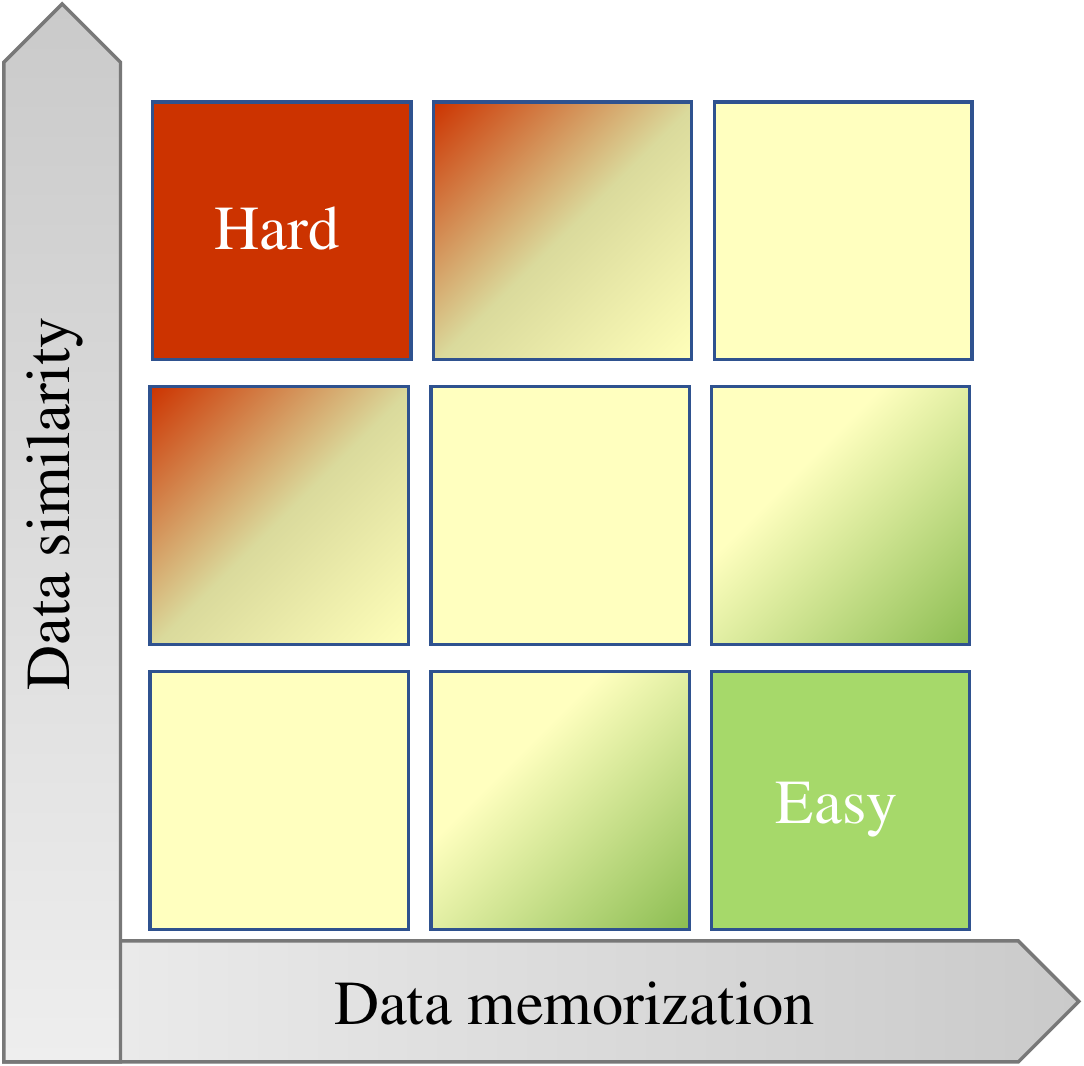}}
  \hspace{.1in}
  \caption{(a) The differential mechanism on different epochs. (b) The relation of data similarity, data memorization and ADP problem.}
 \label{fig:PA}
\end{figure}


(1) \textbf{Auditing Data Distribution.} One challenge of the auditing problem is how to effectively handle the situation when there exists a high degree of similarity between training and non-training data points. Intuitively, it is hard to discriminate training and non-training data points if the two are highly similar. Table \ref{Tab:relation of ds and dm} depicts the relation between data similarity and differential mechanism. The table shows results on CIFAR10 data set with the target model set as GoogleNet. For each class, we calculate the average pair-wise Euclidean distance between training and non-training data points, i.e., the smaller the distance value, the more similar the two data sets. For class `frog', there is a high degree of similarity between training and non-training data points (as shown with the lowest value of 54.21). In this case, while the accuracy achieves 80\%, the differential mechanism generates the worst result, i.e., it corresponds to the largest p-value. These data indicate that the high degree of similarity poses a big challenge to \textsf{DPDA}.

We put forward a plan to address this issue as our future work -- auditing functions can be treated as a transfer function in that it shifts the original data space to a new space where it is easier to solve the problem of the high degree of similarity. Alternatively, a method like Generative Adversarial Networks \cite{DBLP:conf/nips/GoodfellowPMXWOCB14} can be used to achieve this transfer, which offers a better capture and understanding of the distribution of training and non-training data.



(2) \textbf{Data Memorization.} Data memorization refers to a model's capacity to remember its input data, especially its training data \cite{DBLP:conf/icml/ArpitJBKBKMFCBL17}. In \textsf{DPDA}, data memorization of the target model is another important factor for the effectiveness of the differential mechanism. Figure \ref{fig:PA} (a) plots the differential mechanism along different training epochs. It can be observed that the performance of target model is often poor due to under-fitting in early epochs of model training.  It is fair to say that the target model has poor memorization on training data during this period. It follows that the difference between the model output of training and non-training is not sufficiently significant, resulting in larger p-values. However, it can be clearly observed that, as the number of epochs increases, the target model's memorization of training data gets better and better, resulting in smaller and smaller p-values. This demonstrates that, the better the data memorization of the target model, the more effective the proposed differential mechanism for the \emph{ADP} problem.




In conclusion, we describe the relation of data similarity, target model memorization and differential mechanism in \ref{fig:PA} (b). In face of data with high similarity and model with poor memorization, it is hard for the differential mechanism to handle the \emph{ADP} problem. In contrast, for data with low similarity and model with good memorization, the differential mechanism would work well.



\section{conclusions}
\label{sec_conclusion}


This paper investigates an important problem in data provenance, i.e., algorithmically check if a piece of data has been used to train a machine learning model. We introduce a new auditing framework \textsf{DPDA} based on the idea of differential and propose two implementations of the auditing function, additive implementation and multiplicative implementation. Extensive  experiments on real-world data sets have demonstrated the effectiveness of both the proposed methods. We provide discussions for some important aspects of our framework in the ADP problem.




\bibliographystyle{IEEEtran}
\bibliography{IEEEabrv}

\begin{thebibliography}{10}
\providecommand{\url}[1]{#1}
\csname url@samestyle\endcsname
\providecommand{\newblock}{\relax}
\providecommand{\bibinfo}[2]{#2}
\providecommand{\BIBentrySTDinterwordspacing}{\spaceskip=0pt\relax}
\providecommand{\BIBentryALTinterwordstretchfactor}{4}
\providecommand{\BIBentryALTinterwordspacing}{\spaceskip=\fontdimen2\font plus
\BIBentryALTinterwordstretchfactor\fontdimen3\font minus
  \fontdimen4\font\relax}
\providecommand{\BIBforeignlanguage}[2]{{%
\expandafter\ifx\csname l@#1\endcsname\relax
\typeout{** WARNING: IEEEtran.bst: No hyphenation pattern has been}%
\typeout{** loaded for the language `#1'. Using the pattern for}%
\typeout{** the default language instead.}%
\else
\language=\csname l@#1\endcsname
\fi
#2}}
\providecommand{\BIBdecl}{\relax}
\BIBdecl

\bibitem{DBLP:conf/icdt/BunemanKT01}
P.~Buneman, S.~Khanna, and W.~C. Tan, ``Why and where: {A} characterization of
  data provenance,'' in \emph{ICDT}, 2001, pp. 316--330.

\bibitem{DBLP:conf/sp/ShokriSSS17}
R.~Shokri, M.~Stronati, C.~Song, and V.~Shmatikov, ``Membership inference
  attacks against machine learning models,'' in \emph{S\&P}, 2017, pp. 3--18.

\bibitem{DBLP:conf/kdd/SongS19}
C.~Song and V.~Shmatikov, ``Auditing data provenance in text-generation
  models,'' in \emph{KDD}, 2019, pp. 196--206.

\bibitem{DBLP:journals/tdp/RahmanRLM18}
M.~A. Rahman, T.~Rahman, R.~Lagani{\`{e}}re, and N.~Mohammed, ``Membership
  inference attack against differentially private deep learning model,''
  \emph{Transactions on Data Privacy}, vol.~11, no.~1, pp. 61--79, 2018.

\bibitem{yang2020federated}
Q.~Yang, \emph{Federated Learning: Privacy and Incentive}.\hskip 1em plus 0.5em
  minus 0.4em\relax Springer Nature, 2020.

\bibitem{DBLP:conf/ccs/SongRS17}
C.~Song, T.~Ristenpart, and V.~Shmatikov, ``Machine learning models that
  remember too much,'' in \emph{CCS}, 2017, pp. 587--601.

\bibitem{DBLP:conf/sp/NasrSH19}
M.~Nasr, R.~Shokri, and A.~Houmansadr, ``Comprehensive privacy analysis of deep
  learning: Passive and active white-box inference attacks against centralized
  and federated learning,'' in \emph{S\&P}, 2019, pp. 739--753.

\bibitem{DBLP:conf/ndss/Salem0HBF019}
A.~Salem, Y.~Zhang, M.~Humbert, P.~Berrang, M.~Fritz, and M.~Backes,
  ``Ml-leaks: Model and data independent membership inference attacks and
  defenses on machine learning models,'' in \emph{NDSS}, 2019.

\bibitem{shejwalkar2021membership}
V.~Shejwalkar, H.~A. Inan, A.~Houmansadr, and R.~Sim, ``Membership inference
  attacks against {NLP} classification models,'' in \emph{NeurIPS 2021 Workshop
  Privacy in Machine Learning}, 2021.

\bibitem{DBLP:journals/corr/abs-2103-07853}
H.~Hu, Z.~Salcic, G.~Dobbie, and X.~Zhang, ``Membership inference attacks on
  machine learning: {A} survey,'' \emph{CoRR}, vol. abs/2103.07853, 2021.

\bibitem{DBLP:journals/corr/abs-1807-09173}
S.~Truex, L.~Liu, M.~E. Gursoy, L.~Yu, and W.~Wei, ``Towards demystifying
  membership inference attacks,'' \emph{CoRR}, vol. abs/1807.09173, 2018.

\bibitem{DBLP:conf/icml/Choquette-ChooT21}
C.~A. Choquette{-}Choo, F.~Tram{\`{e}}r, N.~Carlini, and N.~Papernot,
  ``Label-only membership inference attacks,'' in \emph{ICML}, 2021, pp.
  1964--1974.

\bibitem{DBLP:conf/sp/PanZJY20}
X.~Pan, M.~Zhang, S.~Ji, and M.~Yang, ``Privacy risks of general-purpose
  language models,'' in \emph{S\&P}, 2020, pp. 1314--1331.

\bibitem{10.1145/3460120.3484770}
M.~Zhang, Z.~Ren, Z.~Wang, P.~Ren, Z.~Chen, P.~Hu, and Y.~Zhang, ``Membership
  inference attacks against recommender systems,'' in \emph{CCS}, 2021, p.
  864–879.

\bibitem{DBLP:conf/kdd/Pei20}
J.~Pei, ``Data pricing - from economics to data science,'' in \emph{KDD}, 2020,
  pp. 3553--3554.

\bibitem{DBLP:journals/popets/HayesMDC19}
J.~Hayes, L.~Melis, G.~Danezis, and E.~D. Cristofaro, ``{LOGAN:} membership
  inference attacks against generative models,'' \emph{Proceedings on Privacy
  Enhancing Technologies}, vol. 2019, no.~1, pp. 133--152, 2019.

\bibitem{DBLP:journals/corr/abs-1802-04889}
Y.~Long, V.~Bindschaedler, L.~Wang, D.~Bu, X.~Wang, H.~Tang, C.~A. Gunter, and
  K.~Chen, ``Understanding membership inferences on well-generalized learning
  models,'' \emph{CoRR}, vol. abs/1802.04889, 2018.

\bibitem{DBLP:conf/ccs/SongR20}
C.~Song and A.~Raghunathan, ``Information leakage in embedding models,'' in
  \emph{CCS}, 2020, pp. 377--390.

\bibitem{DBLP:conf/iclr/ZhangBHRV17}
C.~Zhang, S.~Bengio, M.~Hardt, B.~Recht, and O.~Vinyals, ``Understanding deep
  learning requires rethinking generalization,'' in \emph{ICLR}, 2017.

\bibitem{DBLP:conf/uss/Carlini0EKS19}
N.~Carlini, C.~Liu, {\'{U}}.~Erlingsson, J.~Kos, and D.~Song, ``The secret
  sharer: Evaluating and testing unintended memorization in neural networks,''
  in \emph{USENIX Association}, 2019, pp. 267--284.

\bibitem{DBLP:conf/icalp/Dwork06}
C.~Dwork, ``Differential privacy,'' in \emph{ICALP}, 2006.

\bibitem{DBLP:journals/corr/JiLE14}
Z.~Ji, Z.~C. Lipton, and C.~Elkan, ``Differential privacy and machine learning:
  a survey and review,'' \emph{CoRR}, vol. abs/1412.7584, 2014.

\bibitem{DBLP:conf/webi/VaidyaSBH13}
J.~Vaidya, B.~Shafiq, A.~Basu, and Y.~Hong, ``Differentially private naive
  bayes classification,'' in \emph{ICWI}, 2013, pp. 571--576.

\bibitem{DBLP:conf/nips/ChaudhuriSS12}
K.~Chaudhuri, A.~D. Sarwate, and K.~Sinha, ``Near-optimal differentially
  private principal components,'' in \emph{NIPS}, 2012, pp. 998--1006.

\bibitem{DBLP:journals/jpc/RubinsteinBHT12}
B.~I.~P. Rubinstein, P.~L. Bartlett, L.~Huang, and N.~Taft, ``Learning in a
  large function space: Privacy-preserving mechanisms for {SVM} learning,''
  \emph{Journal of Privacy and Confidentiality}, vol.~4, no.~1, 2012.

\bibitem{DBLP:conf/icdm/JagannathanPW09}
G.~Jagannathan, K.~Pillaipakkamnatt, and R.~N. Wright, ``A practical
  differentially private random decision tree classifier,'' in \emph{ICDM
  Workshops}, 2009, pp. 114--121.

\bibitem{DBLP:conf/ccs/AbadiCGMMT016}
M.~Abadi, A.~Chu, I.~J. Goodfellow, H.~B. McMahan, I.~Mironov, K.~Talwar, and
  L.~Zhang, ``Deep learning with differential privacy,'' in \emph{CCS}, 2016,
  pp. 308--318.

\bibitem{DBLP:conf/ccs/ShokriS15}
R.~Shokri and V.~Shmatikov, ``Privacy-preserving deep learning,'' in
  \emph{CCS}, 2015, pp. 1310--1321.

\bibitem{DBLP:journals/corr/abs-1912-04977}
P.~Kairouz, H.~B. McMahan \emph{et~al.}, ``Advances and open problems in
  federated learning,'' \emph{CoRR}, vol. abs/1912.04977, 2019.

\bibitem{DBLP:journals/corr/abs-1712-07557}
R.~C. Geyer, T.~Klein, and M.~Nabi, ``Differentially private federated
  learning: {A} client level perspective,'' \emph{CoRR}, vol. abs/1712.07557,
  2017.

\bibitem{DBLP:books/lib/MichalskiA84}
R.~S. Michalski and J.~R. Anderson, \emph{Machine learning - an artificial
  intelligence approach}, ser. Symbolic computation.\hskip 1em plus 0.5em minus
  0.4em\relax Springer, 1984.

\bibitem{DBLP:journals/ml/AhaKA91}
D.~W. Aha, D.~F. Kibler, and M.~K. Albert, ``Instance-based learning
  algorithms,'' \emph{Machine Learning}, vol.~6, pp. 37--66, 1991.

\bibitem{DBLP:conf/iclr/NovakBAPS18}
R.~Novak, Y.~Bahri, D.~A. Abolafia, J.~Pennington, and J.~Sohl{-}Dickstein,
  ``Sensitivity and generalization in neural networks: an empirical study,'' in
  \emph{ICLR}, 2018.

\bibitem{DBLP:conf/aaai/ShuZ19}
H.~Shu and H.~Zhu, ``Sensitivity analysis of deep neural networks,'' in
  \emph{AAAI}, 2019, pp. 4943--4950.

\bibitem{DBLP:conf/icpr/ForouzeshST20}
M.~Forouzesh, F.~Salehi, and P.~Thiran, ``Generalization comparison of deep
  neural networks via output sensitivity,'' in \emph{ICPR}, 2020, pp.
  7411--7418.

\bibitem{DBLP:conf/cvpr/SzegedyLJSRAEVR15}
C.~Szegedy, W.~Liu, Y.~Jia, P.~Sermanet, S.~E. Reed, D.~Anguelov, D.~Erhan,
  V.~Vanhoucke, and A.~Rabinovich, ``Going deeper with convolutions,'' in
  \emph{CVPR}, 2015, pp. 1--9.

\bibitem{DBLP:conf/nips/KrizhevskySH12}
A.~Krizhevsky, I.~Sutskever, and G.~E. Hinton, ``Imagenet classification with
  deep convolutional neural networks,'' in \emph{NIPS}, 2012, pp. 1106--1114.

\bibitem{DBLP:conf/ijcai/ZhouF17}
Z.~Zhou and J.~Feng, ``Deep forest: Towards an alternative to deep neural
  networks,'' in \emph{IJCAI}, 2017, pp. 3553--3559.

\bibitem{DBLP:journals/corr/abs-1811-03378}
C.~Nwankpa, W.~Ijomah, A.~Gachagan, and S.~Marshall, ``Activation functions:
  Comparison of trends in practice and research for deep learning,''
  \emph{CoRR}, vol. abs/1811.03378, 2018.

\bibitem{DBLP:journals/corr/abs-1810-00069}
A.~Chakraborty, M.~Alam, V.~Dey, A.~Chattopadhyay, and D.~Mukhopadhyay,
  ``Adversarial attacks and defences: {A} survey,'' \emph{CoRR}, vol.
  abs/1810.00069, 2018.

\bibitem{DBLP:conf/iclr/KurakinGB17a}
A.~Kurakin, I.~J. Goodfellow, and S.~Bengio, ``Adversarial examples in the
  physical world,'' in \emph{ICLR}, 2017.

\bibitem{DBLP:conf/iclr/KurakinGB17}
------, ``Adversarial machine learning at scale,'' in \emph{ICLR}, 2017.

\bibitem{DBLP:journals/corr/GoodfellowSS14}
I.~J. Goodfellow, J.~Shlens, and C.~Szegedy, ``Explaining and harnessing
  adversarial examples,'' in \emph{ICLR}, 2015.

\bibitem{DBLP:journals/tkde/MuTZ17}
X.~Mu, K.~M. Ting, and Z.~Zhou, ``Classification under streaming emerging new
  classes: {A} solution using completely-random trees,'' \emph{{IEEE}
  Transactions on Knowledge and Data Engineering}, vol.~29, no.~8, pp.
  1605--1618, 2017.

\bibitem{DBLP:journals/access/ZhangGGSYHZL17}
K.~Zhang, C.~Gao, L.~Guo, M.~Sun, X.~Yuan, T.~X. Han, Z.~Zhao, and B.~Li, ``Age
  group and gender estimation in the wild with deep ror architecture,''
  \emph{{IEEE} Access}, vol.~5, pp. 22\,492--22\,503, 2017.

\bibitem{DBLP:conf/bigdataconf/SmithC18}
P.~Smith and C.~Chen, ``Transfer learning with deep cnns for gender recognition
  and age estimation,'' in \emph{{IEEE} BigData}, 2018, pp. 2564--2571.

\bibitem{DBLP:journals/tifs/EidingerEH14}
E.~Eidinger, R.~Enbar, and T.~Hassner, ``Age and gender estimation of
  unfiltered faces,'' \emph{{IEEE} Transactions on Information Forensics and
  Security}, vol.~9, no.~12, pp. 2170--2179, 2014.

\bibitem{DBLP:conf/cvpr/HeZRS16}
K.~He, X.~Zhang, S.~Ren, and J.~Sun, ``Deep residual learning for image
  recognition,'' in \emph{CVPR}, 2016, pp. 770--778.

\bibitem{DBLP:conf/bmvc/ParkhiVZ15}
O.~M. Parkhi, A.~Vedaldi, and A.~Zisserman, ``Deep face recognition,'' in
  \emph{British Machine Vision Conference}, 2015, pp. 41.1--41.12.

\bibitem{DBLP:conf/nips/GoodfellowPMXWOCB14}
I.~J. Goodfellow, J.~Pouget{-}Abadie, M.~Mirza, B.~Xu, D.~Warde{-}Farley,
  S.~Ozair, A.~C. Courville, and Y.~Bengio, ``Generative adversarial nets,'' in
  \emph{NIPS}, 2014, pp. 2672--2680.

\bibitem{DBLP:conf/icml/ArpitJBKBKMFCBL17}
D.~Arpit, S.~Jastrzebski, N.~Ballas, D.~Krueger, E.~Bengio, M.~S. Kanwal,
  T.~Maharaj, A.~Fischer, A.~C. Courville, Y.~Bengio, and S.~Lacoste{-}Julien,
  ``A closer look at memorization in deep networks,'' in \emph{ICML}, 2017, pp.
  233--242.

\end{thebibliography}



\section{Supplementary Materials}

\subsection{Experiment Setup}

{\bf Data Sets. } We use 
four datasets to compare the performance of all methods: 

(1) {\bf MNIST\footnote{http://yann.lecun.com/exdb/mnist/}}: The MNIST is an image dataset of handwritten digits. It contians 10 classes and 784 features; 

(2) {\bf 20 Newsgroups\footnote{http://qwone.com/~jason/20Newsgroups/}}: The dataset collates approximately 20,000 newsgroup documents partitioned across 20 different newsgroups. The Word2vec is used to prepossess text data. 

(3) {\bf CIFAR-10\footnote{https://www.cs.toronto.edu/~kriz/cifar.html}}: It is a standard classification dataset consisting of 32$\times$32 color images belonging to 10 different object classes.

(4) {\bf VGGFace\footnote{https://www.robots.ox.ac.uk/~vgg/data/vgg\_face/}}: The dataset consists of the crawled images of celebrities on the Web. There are 2622 celebrities in the dataset.

\noindent{\bf Evaluation Metrics.} We use \textbf{F-measure} to measure the performance. This measure produces a combined effect of precision (P) and recall (R) of the auditing performance, 
\[ F\texttt{-}measure = \frac{2*P*R}{P+R}.\] 
F-measure = 1 if the method identifies all training data with no false positives. 

We also employ \textbf{AUC}\footnote{https://en.wikipedia.org/wiki/Receiver\_operating\_characteristic} (``Area under the ROC Curve'') to access performance. An ROC curve (receiver operating characteristic curve) is a graph showing the performance of a classification model at all classification thresholds. This curve plots True Positive Rate (TPR) and False Positive Rate (FPR).

\subsection{Results on synthetic data} 


\subsubsection{SVM-based model} SVM is a discriminative classifier which classifies new data points by calculating an optimal separating hyperplane\footnote{Corinna Cortes and Vladimir Vapnik. 1995. Support-Vector Networks. Machine Learning 20, 3 (1995), 273–297.}.  In two dimensional space this hyperplane is a line dividing a plane into two parts each defining a class for data points within it. In this paper, we illustrate with a least squares SVM classifier\footnote{Johan A. K. Suykens and Joos Vandewalle. 1999. Least Squares Support VectorMachine Classifiers.Neural Processing Letters 9, 3 (1999), 293–300.}. Given a set of instance-label pairs $(x_i, y_i)$, $i = 1,\cdots,l$, $x_i \in \mathbb{R}^d$, $y_i \in \{-1, +1\}$, it solves the following optimization problem:
 $\min_w \ \frac{1}{2} w^{T}w +C\sum_{i=1}^{l} \xi^2 $, 
subject to the equality constraints: $y_i(w x_i+b) = 1 - \xi_i$, where $C > 0$ is a penalty parameter. We set the partial derivatives of $x$ on the cost:
\begin{equation*}
\begin{aligned}
\frac{\partial J(w)}{\partial x}= 2[1-wx)]w.
\end{aligned}
\end{equation*}
Then, we can calculate $\eta$ by Eqn.(\ref{section5-eta}).


\subsubsection{Tree-based model} We discuss random forest model with completely-random trees as target model in this part\footnote{Xindong Wu, Vipin Kumar, J. Ross Quinlan, Joydeep Ghosh, et. al. 2008. Top 10 algorithms in data mining. Knowledge and Information Systems 14, 1 (2008), 1–37.}. random forest model is usually trained with the ``bagging'' method. The general idea of the bagging method is that a combination of learning models increases the overall result. Each tree in the classifications takes input from samples in the initial dataset. Features are then randomly selected, which are used in growing the tree at each node. Every tree in the forest should not be pruned until the end of the exercise when the prediction is reached decisively.

\begin{figure*}[t]
  \centering
\subfigure[Auditing data.]{\includegraphics[height=1.5in,width=1.7in]{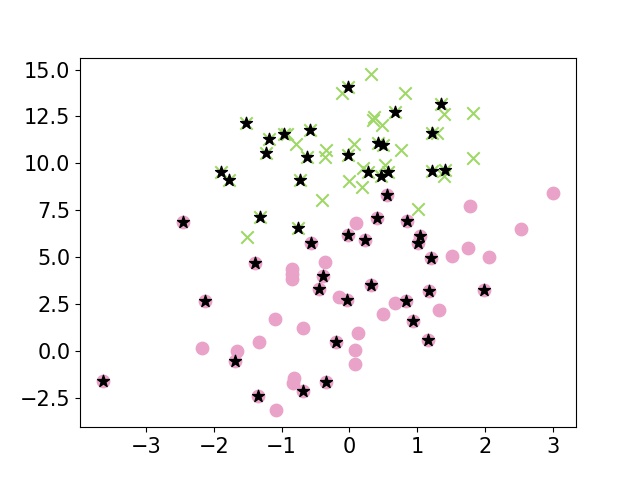}}
 \hspace{.2in}
\subfigure[Results on white-box condition.]{\includegraphics[height=1.5in,width=1.7in]{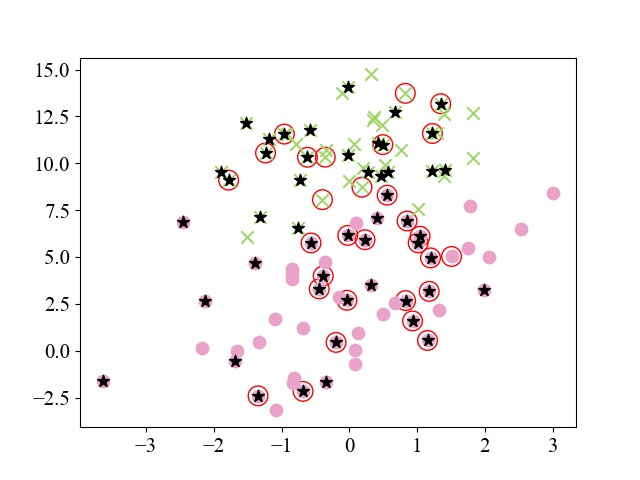}}
 \hspace{.2in}
\subfigure[Results on black-box condition.]{\includegraphics[height=1.5in,width=1.7in]{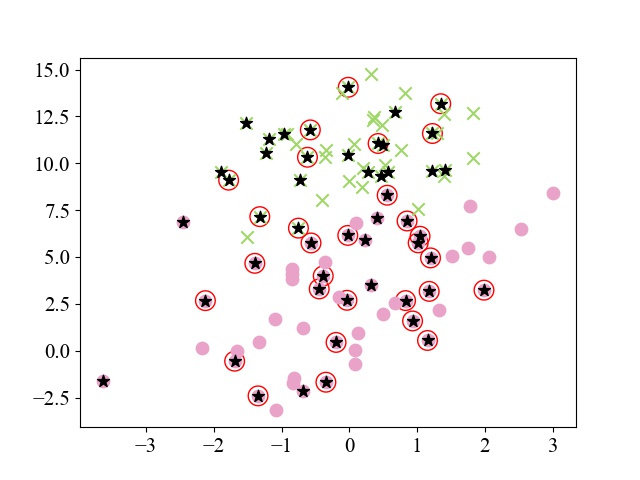}}
  \caption{An illustration on the Tree-based model. Training data are marked by black star marks training data. Red circle is the auditing results. (b) is result under white-box condition, (c) is result under black-box condition.
}  
\label{example44}
\end{figure*}

Note that because the random forest is a non-linear form, it is not easy to use the gradient to calculate $\eta$. In this paper, we attempt to set Gaussian noise perturbation as $\eta$ to observe this result. Figure \ref{example44} shows the complete illustrations.

\subsubsection{Neural network-baesd} A neural network (NN) is a technique that uses a hierarchical composition of $n$ parametric functions to model an input $x$. Each function $f_i$ for $i \in 1,\cdots,n$ is modeled using a layer of neurons, which are elementary computing units applying an activation function to the previous layer’s weighted representation of the input to generate a new representation. Each layer is parameterized by a weight vector $\theta_i$ impacting each neuron’s activation. Such weights hold the knowledge of a NN model and are evaluated during its training phase, as detailed below. Thus, a NN defines and computes:
\begin{equation}
    M(x) = f_n (\theta_n, f_{n-1} (\theta_{n-1}, \cdots f_2 (\theta_2, f_1 (\theta_1, x)))) 
\end{equation}
At each layer: $y_j = f(\sum^{3}_{i=1} w_{ij} x_i + b)$, where $W_{ij}$, $x_i$ and $y_j$ are the weights, input and output respectively. We show a two-linear-layer NN example as follow:  
\begin{equation}
\begin{aligned}
    M(x) &= w_{2}(w_{1}  x + b_1)+ b_2 \\
        &  = w_{2}w_{1} x + w_{2}b_1+ b_2
\end{aligned}
\end{equation}

\begin{figure*}[t]
  \centering
\subfigure[Auditing data and hyperplane.]{\includegraphics[height=1.35in,width=1.55in]{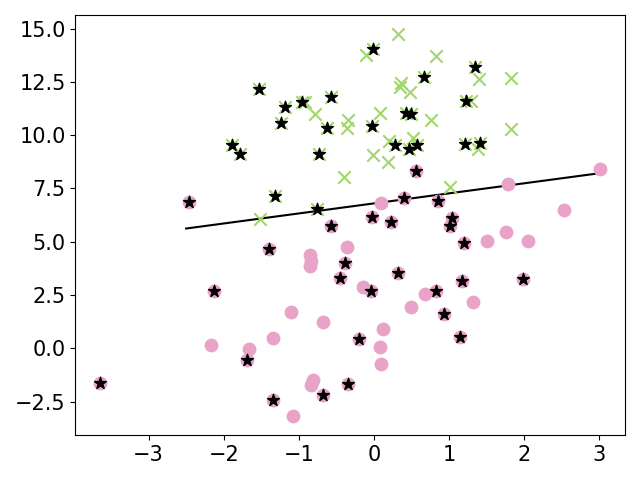}}
 \hspace{.2in}
\subfigure[Results on white-box condition.]{\includegraphics[height=1.5in,width=1.7in]{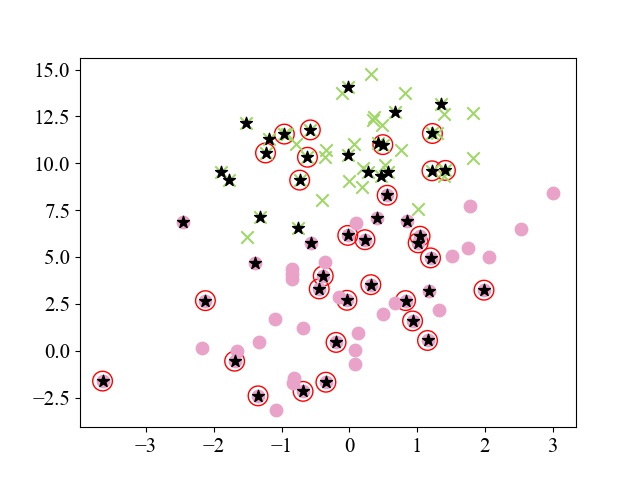}}
 \hspace{.2in}
\subfigure[Results on black-box condition.]{\includegraphics[height=1.35in,width=1.55in]{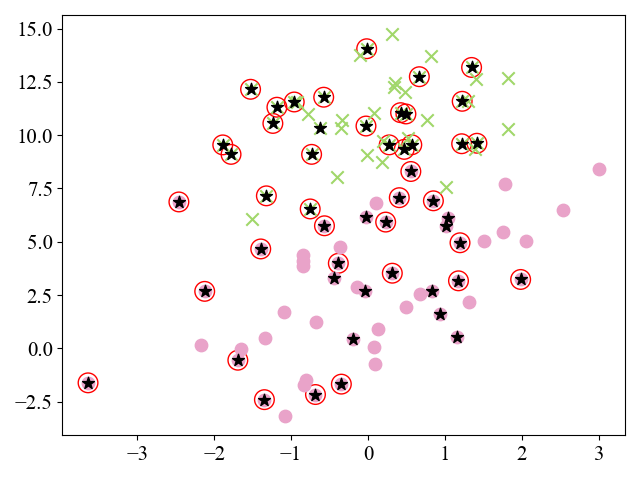}}
 \caption{An illustration on the NN-based model. Training data are marked by black star marks training data. Red circle is the auditing results. (b) is result under white-box condition, (c) is result on black-box condition.
}  
\label{example77}
\end{figure*}

The above equation can be view as an SVM-based model. We, therefore, apply the same way to calculate $\eta$. Figure \ref{example77} shows NN-based model results.

\begin{figure*}[t]
  \centering
\subfigure[Results on SVM-based model.]{\includegraphics[height=1.5in,width=1.7in]{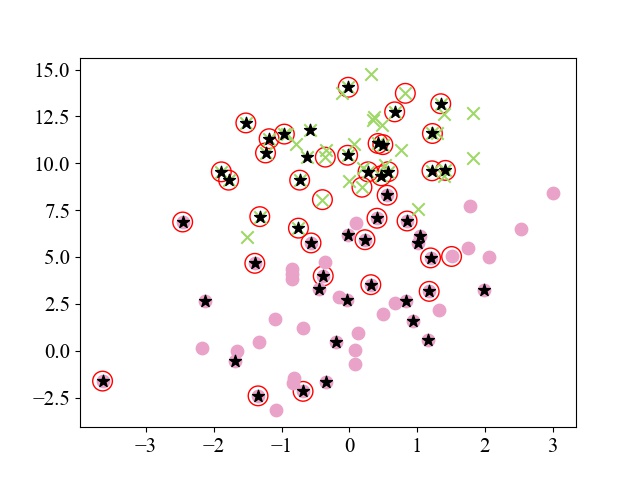}}
 \hspace{.2in}
\subfigure[Results on Tree-based model.]{\includegraphics[height=1.5in,width=1.7in]{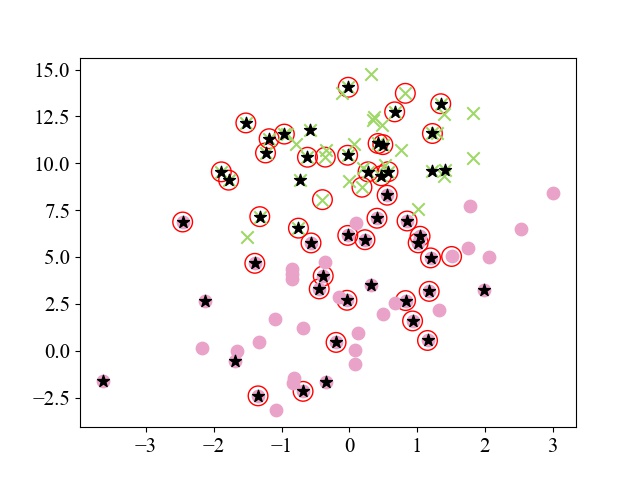}}
 \hspace{.2in}
\subfigure[Results on NN-based model.]{\includegraphics[height=1.5in,width=1.7in]{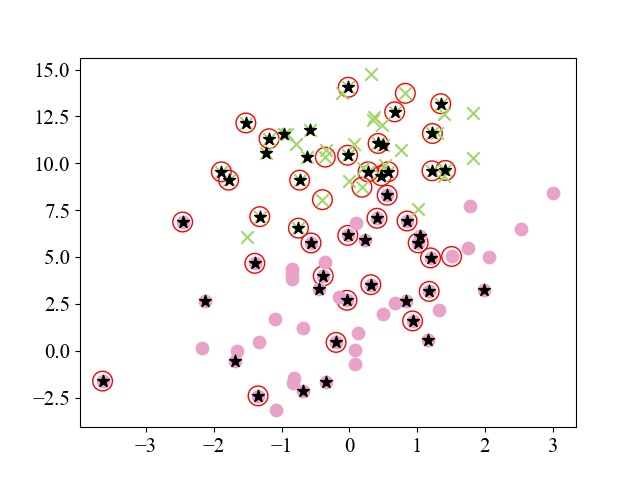}}
  \caption{An illustration on the results of multiplicative implementation. Black star marks training data. red circle marks auditing results. (a) is SVM-based model result, (b) is tree-based model result, (c) is NN-based model result.
}  
\label{examplemapping}
\end{figure*}

Figure \ref{examplemapping} shows the results of multiplicative implementation on synthetic data data, the experiment setup is the same as Section \ref{Illustration on synthetic data}. (a)-(c) are results under black-box condition. 

\end{document}